\documentclass[12pt]{article}

\usepackage{multirow}
\usepackage[margin = 2cm]{geometry}
\usepackage[center]{caption}
\usepackage{tablefootnote}
\usepackage{amsmath,amsfonts,amssymb}
 \usepackage{amsmath} 
\usepackage[utf8]{inputenc} 
\usepackage[T1]{fontenc}    
\usepackage{hyperref}       
\usepackage{url}            
\usepackage{booktabs}       
\usepackage{amsfonts}       
\usepackage{nicefrac}       
\usepackage{microtype}      
\usepackage{xcolor}         
\usepackage{graphicx}
\usepackage{float}
\usepackage{caption}
\usepackage{subcaption}
\usepackage{appendix}
\usepackage[version=4]{mhchem}

\usepackage{ccicons}

\DeclareMathOperator*{\concat}{%
    \mathchoice%
        {\Big\Vert}%
        {\big\Vert}%
        {\Vert}%
        {\Vert}%
}

\DeclareMathOperator*{\argminA}{arg\,max} 

\usepackage{breakurl}
\usepackage{hyperref}
\hypersetup{breaklinks=true}

\usepackage{authblk}
\usepackage[natbibapa]{apacite}
\usepackage{caption}
\begin{document}

\title{INFLECT-DGNN: Influencer Prediction with Dynamic Graph Neural Networks\footnote{\scriptsize NOTICE: This is a preprint of a published work. Changes resulting from the publishing process, such as editing, corrections,structural formatting, and other quality control mechanisms may not be reflected in this version of the document. Please cite this work as follows: E. Tiukhova, E. Penaloza, M. Óskarsdóttir, B. Baesens, M. Snoeck and C. Bravo, "INFLECT-DGNN: Influencer Prediction With Dynamic Graph Neural Networks," in IEEE Access, vol. 12, pp. 115026-115041, 2024, DOI: \href{https://doi.org/10.1109/ACCESS.2024.3443533}{10.1109/ACCESS.2024.3443533}. This work is made available under a \href{https://creativecommons.org/licenses/by/4.0/}{Creative Commons BY license}. \ccby}}

\author[1]{Elena Tiukhova}
\author[2]{Emiliano Penaloza}
\author[3]{María Óskarsdóttir}
\author[1,4]{Bart Baesens}
\author[1]{Monique Snoeck}
\author[2]{Cristi\'{a}n Bravo}

\affil[1]{Research Center for Information Systems Engineering, Faculty of Economics and Business, KU Leuven, Naamsesstraat 69, 3000, Leuven, Belgium}
\affil[2]{Department of Statistical and Actuarial Sciences, Western University, 1151 Richmond Street, London, Ontario, N6A 5B7, Canada}
\affil[3]{Department of Computer Science, Reykjavík University, Menntavegi 1, 102 Reykjavík, Iceland}
\affil[4]{Department of Decision Analytics and Risk, University of Southampton, University Road, Southampton, SO17 1BJ, UK}

\date{}
\maketitle
\begin{abstract}
Leveraging network information for predictive modeling has become widespread in many domains. Within the realm of referral and targeted marketing, influencer detection stands out as an area that could greatly benefit from the incorporation of dynamic network representation due to the continuous evolution of customer-brand relationships. In this paper, we present INFLECT-DGNN, a new method for profit-driven INFLuencer prEdiCTion with Dynamic Graph Neural Networks that innovatively combines Graph Neural Networks (GNNs) and Recurrent Neural Networks (RNNs) with weighted loss functions, synthetic minority oversampling adapted to graph data, and a carefully crafted rolling-window strategy. We introduce a novel profit-driven framework that supports decision-making based on model predictions. To test the framework, we use a unique corporate dataset with diverse networks, capturing the customer interactions across three cities with different socioeconomic and demographic characteristics. Our results show how using RNNs to encode temporal attributes alongside GNNs significantly improves predictive performance, while the profit-driven framework determines the optimal classification threshold for profit maximization. We compare the results of different models to demonstrate the importance of capturing network representation, temporal dependencies, and using a profit-driven evaluation. Our research has significant implications for the fields of referral and targeted marketing, expanding the technical use of deep graph learning within corporate environments.
\end{abstract}

\begin{keywords}
Dynamic Graph Neural Networks, Graph Isomorphism Networks (GINs), Graph Attention Networks (GATs), referral marketing, influencer prediction
\end{keywords}

\section{Introduction}
\label{Introduction}

Advances in analytics have enabled organizations to leverage data for effective decision-making and increased value creation \citep{hussain2024magres}. A growing number of social interactions have become a valuable source for constructing networks that provide useful insights into interconnectivity patterns between individuals and can serve as a data source for real-world applications such as fraud detection \citep{oskarsdottir2022social}, recommender systems \citep{yang2023heterogeneous}, fake news detection \citep{sivasankari2022tracing} and marketing \citep{rani2022influential}. Word of Mouth (WOM) has long been recognized as a powerful tool for customer influence in marketing to spread information effectively in customer networks \citep{puigbo2014influencer}. Customers who are able to persuade others within their network can be considered influencers or opinion leaders \citep{puigbo2014influencer}. Identifying and incentivizing them early helps to engage more customers in profitable initiatives by effectively spreading information across the network.

Typically, the technique used to incentivize customers who are considered influencers falls within the context of referral marketing, which leverages social connections to promote a product or service \citep{roelens2018advances}. In this context, the topology of social networks is considered a major factor in the dissemination of information, with influencers being the best candidates to make referrals. To harness the power of social networks, it is essential to go beyond simply identifying opinion leaders. Anticipating the rise of potential influencers prior to their network enrollment, while also enabling re-influencing actions from past influencers who could repeatedly refer new customers, can serve as a significant competitive accelerator. Moreover, the value of predictive models depends heavily on the profit generated when the model's predictions are used in real marketing campaigns. Cost-sensitive decision making based on instance-dependent threshold tuning has proven to result in the highest savings \citep{vanderschueren2022predict}. This is achieved by complimeting a cost-insensitive classifier with cost-sensitive threshold optimization, which is known as the ``predict-then-optimize'' approach \citep{vanderschueren2022predict}. As a result, the use of influencer prediction models should not only be based on predictive performance, but also integrate a business view and evaluate the potential profit of the action, a property that is often overlooked in the existing studies. By doing so, businesses can effectively leverage the power of social networks to both expand their reach and drive financial growth, as well as more effectively target customers, reducing unnecessary spam and providing both a more nuanced approach to customer incentivizing and a better customer experience.

Network data is non-linear because it cannot be readily represented in Euclidean space. Therefore, the network topology should be extracted first to be used for downstream tasks. There are standard approaches for encoding network topology, such as neighborhood and centrality metrics, as well as collective inference algorithms \citep{baesens2015fraud}. With advances in computing power and deep learning, Graph Neural Networks (GNNs) have become a popular neural network architecture for learning from graph data \citep{zhou2020graph}. GNNs have been successfully leveraged in different domains, including influencer detection \citep{zheng2020demand,shi2022graph}. To account for dynamic networks, i.e. networks that change over time, Dynamic GNNs (DGNNs) came to the forefront of network analysis \citep{skarding2021foundations}. \cite{oskarsdottir2018time} showed that incorporating the temporal component of the network improves predictive performance. Therefore, GNNs can be used as the primary network representation learning technique and further complemented with the temporal component (e.g., recurrent neural networks (RNNs)) to account for the dynamic nature of social networks.

This paper presents INFLECT-DGNN, a novel method designed for influencer prediction in growing customer networks in a profit-driven manner. By leveraging the power of DGNNs to effectively learn from evolving network data, we solve a complex influencer prediction problem in dynamic networks with multiple relationships using edge coloring, i.e., attributing specific characteristics to edges. Our method integrates two types of GNNs (Graph Attention Network (GAT) and Graph Isomorphism Network (GIN)) and Recurrent Neural Networks (RNNs). To learn from imbalanced data, INFLECT incorporates weighted loss functions and the Synthetic Minority Oversampling TEchnique (SMOTE) adapted for graphs \citep{zhao2021graphsmote}. INFLECT-DGNN also introduces a new profit-driven formula to enhance its practical applicability for decision making. Hence, we make the following \textbf{contributions}: 

\begin{enumerate}
    
    \item We present a new method for constructing dynamic attributed edge-colored customer networks with a labeling intuition that allows for re-influencing prediction. 
    \item We develop DGNN models by combining GNNs and RNNs, i.e., GATs and GINs with Long-Short Term Memory (LSTM) and Gated Recurrent Unit (GRU) models. We rigorously evaluate the performance of different model combinations to suggest an optimal architecture for deploying our solution. 
    \item We present a new method of inductive learning from imbalanced data by incorporating weighted loss functions, GraphSMOTE, and a rolling-window strategy in a creative way.  
    \item We thoroughly evaluate our method against baseline static GNNs and dynamic non-GNN approaches based on their predictive performance as well as ecological footprint.\footnote{The source code and online appendices can be found at \url{https://github.com/Banking-Analytics-Lab/INFLECT}}
    \item We introduce a new profit framework for influencer prediction to decide on the optimal classification threshold and evaluate models in a profit-driven manner.  
\end{enumerate}

The \textbf{novelty} of our research is that we address influencer prediction on dynamic attributed edge-colored customer networks with cost-sensitive decision making using DGNNs, something that has not been done in prior research.

This paper is organized as follows. Section \ref{related_work} provides an overview of the related studies. In Section \ref{methodology}, the methodology is defined. The experimental setup is outlined in Section \ref{experiments}. We present the results of the experiments in Sections \ref{results} and the profit framework in Section \ref{profit_evaluation}. Section \ref{discussion} discusses the implications of the results. Finally, Section \ref{conclusion} concludes the research by addressing its limitations and suggestions for future research.

\section{Related work}
\label{related_work}

\subsection{Graph Neural Networks}
\label{GNNs}

Deep learning's increasing popularity has brought GNNs to the forefront of representational learning for networks \citep{9046288}. With their versatility, GNNs can solve node-, edge-, and graph-level tasks using supervised, semi-supervised, and unsupervised learning. GNNs can be categorized into several types based on their learning approach, task level, and complexity. Recurrent GNNs, the pioneers in this field, use the same set of parameters recurrently to learn high-level node representations \citep{9046288}. Convolutional GNNs, a more advanced type, generalize the fixed grid structure of Convolutional Neural Networks (CNN) and use either spectral-based or spatial-based graph convolution to aggregate neighbor features and learn node representations \citep{9046288}. Graph Autoencoders learn node embeddings in an unsupervised manner by reconstructing the adjacency matrix, while Spatial–Temporal GNNs enable dynamic network analysis by incorporating the time dimension \citep{9046288}.

Although there are numerous GNNs available, they share a similar design pipeline \citep{zhou2020graph}. It includes the steps of identifying network structure and type, designing the loss function, and building the model using computational modules \citep{zhou2020graph}. Network structure can be either explicit, where the network is given upfront, or implicit, where the network is to be built from the task at hand. The main network types are directed or undirected, homogeneous or heterogeneous, and static or dynamic networks. Loss functions are dependent on the task type, including its level and the type of supervision \citep{zhou2020graph}. Computational modules can be classified into propagation, sampling, and pooling modules. A propagation module is used to propagate feature and topological information between nodes and can include convolution operators, recurrent operators, or skip connections. A sampling module can be utilized to sample large networks, while a pooling module is used to extract information from nodes \citep{zhou2020graph}. 

\subsection{Dynamic Graph Neural Networks}
\label{DyGNNs}
A network that changes over time, with appearing/disappearing nodes and edges, is called a dynamic network \citep{skarding2021foundations}. Research shows that incorporating the temporal component of networks into the prediction task enhances the models' performance \citep{oskarsdottir2018time}.
Dynamic GNNs are often represented using the encoder-decoder structure, where the encoder focuses on learning node embeddings and the decoder is used to solve prediction tasks such as node or edge classification. \cite{zhu2022learnable} examined different encoder-decoder architectures for supervised dynamic network learning, with techniques categorized as either Discrete Time Dynamic Graph (DTDG) or Continuous Time Dynamic Graph (CTDG) learning. The difference between DTDG and CTDG learning is in how they handle time, with DTDG models using network snapshots that capture the topology at a point in time, and CTDG models considering networks as an event stream that updates over time. A further distinction can be made within DTDG based on how structural and temporal patterns are captured by the encoder-decoder composition \citep{skarding2021foundations}. Stacked DTDG models capture structural and temporal information in separate layers, while integrated DTDG models combine them in one layer. The former deals with combining existing layers in new ways (as in \cite{seo2018structured}), while the latter requires a design of new layers (e.g., \cite{pareja2020evolvegcn}). 

\subsection{Referral marketing}
\label{ref_marketing}
Unlike the business-to-customer (B2C) relationships leveraged by traditional marketing techniques, referral marketing leverages customer-to-customer (C2C) relationships to promote products or services.  Referral programs create additional motivation for customers to refer other potential customers in exchange for a reward. The reward can be a coupon, cash, bonus points, bonus products, etc. Targeted and referral marketing programs catalyze WOM effects by reaching influential customers to encourage them to make product recommendations \citep{ryu2007penny}. \cite{ryu2007penny} found that the offer of a reward increases the likelihood of a referral, regardless of the size of the reward offered. In particular, this effect is stronger for weak ties between a referrer and a potential customer, i.e., exchange relationships, as well as for weaker brands. Thus, referrals require more encouragement in the absence of strong ties or high brand recognition. However, even for stronger ties and brands, the reward should be present (non-zero) to reduce the level of inequity for existing customers. Therefore, firms need to balance the marginal revenue impact with the size of the reward and the change in the likelihood of referrals \citep{ryu2007penny}. \cite{ARMELINI2015449} found that referred customers are more loyal than non-referred customers. In addition, they also found that referred customers are more valuable when the referring customer has a high customer lifetime value. This means that referral marketing is most successful when the targeted customers are not only influential, but also profitable. Also, referred customers have a longer customer lifetime than non-referred customers \citep{roelens2018advances}. 

\subsection{Influencer detection in networks}
\label{LR}

Influencer detection has been extensively researched in the social network domain. Valuable insights into a user's network influence can be gleaned from their personal information, platform interactions, and the dynamics of user interactions. Early research focused on different approaches to detect influential network nodes, including network centrality measures, prestige ranking algorithms, and information diffusion methods \citep{puigbo2014influencer}. Influential nodes can also be discovered by following the graph-driven branching process, which generalizes network centrality metrics \citep{de2023branching}. Text data, such as online forums or tweets, can serve as a source for semantic analysis, which has been shown to be useful for filtering out irrelevant network interactions and identifying key nodes using classical authority discovery algorithms \citep{rios2019semantically}. Considering different types of nodes and their interactions can also be advantageous for influencer detection, as in the case of the SNet model \citep{huynh2020detecting}: by examining relationships not only among users, but also between users and posts, influence can be estimated, and detected influencers can be leveraged for influencer marketing. In addition, the scope extends beyond different user categories to heterogeneous social connections that manifest as multi-layered networks. Combined with a heterogeneous degree ranking algorithm, they provide a powerful framework that allows for a better overview of the influential nodes \citep{rani2022influential}.

Influencer detection can also borrow the insights from game theory and deep learning. The former is illustrated by the EDGly model, which goes beyond basic centrality metrics by using the ensemble clustering for graphs to identify important partitioned communities and Shapley value games to uncover influential nodes within these communities \citep{jain2022edgly}. The latter is represented by a growing adoption of GNNs in the influencer detection domain. 

A recent systematic literature review by \cite{seyfosadat2023systematic} outlines the studies on identifying influencers in social networks. Among the machine learning-based and graph-based approaches reviewed by \cite{seyfosadat2023systematic}, the studies that use GNNs to detect influencers are mentioned, including the studies by \cite{zheng2020demand}, \cite{yu2020identifying}, and \cite{zhao2020infgcn}. In addition, Graph Influence Networks have emerged as a specialized GNN technique tailored for influencer discovery by exploiting the inherent local graph structures to identify influential neighbors of a node \citep{shi2022graph}. Table \ref{tab:related_work} provides an overview of the studies mentioned above. 

\begin{table*}[]
\fontsize{8}{10}\selectfont 
\caption{Related work on GNNs for influencer detection}
\begin{tabular}{p{1cm}p{13cm}p{1cm}p{1cm}}
\hline
\textbf{Ref.} & \textbf{Description} & \textbf{GNN used} & \textbf{Dynamic?} \\ \hline
\cite{zheng2020demand} & Network topology and text message semantics are combined for influencer discovery, leveraging both a language attention network (LAN) and a GCN: LAN is used as a subject filter, while GCN is used to model influence propagation. & GCN & No \\
\cite{yu2020identifying} & Inspired by GCN workings, the critical node identification problem in complex networks is transformed into a regression problem: considering both adjacency matrix and convolutional neural networks (CNN), a feature matrix is generated for each node, and CNN is used to train and predict the influence of nodes. & GCN-inspired & No \\
\cite{zhao2020infgcn} & Based on the fixed-size neighbor networks identified by the Breadth First Search algorithm and node features derived from the network (degree, betweenness, closeness and clustering coefficient), the GCN model is used to learn latent representations of the nodes. These representations are then used to solve the downstream prediction task. & GCN & No \\
\cite{shi2022graph} & The most influential nodes in a graph are identified using both topological structures and node features by (1) constructing an influence set of a node based on the local graph structure; (2) measuring the feature influences of the nodes in the influence set; (3) selecting influential neighbors based on structural accessibility and task relevance. & Novel GINN & No \\ \hline
\end{tabular}
\label{tab:related_work}
\end{table*}

\subsection{Research gap}

The current literature reviewed in section \ref{LR} has extensively studied the problem of influencer detection using various economic, statistical and machine learning techniques. However, the majority of these studies consider influencer detection in the online social media platforms and do not cover other domains such as corporate customer networks, which are dynamically changing. Moreover, the research on influencer detection with GNNs is limited because it lacks the incorporation of more powerful and recent techniques.

In our previous work \citep{tiukhova2022influencer}, we introduced an influencer detection task in dynamic customer networks, which are conceptually different from social media networks, and defined influencers as customers who have at least once referred other customers, with the influencing behavior being propagated to the future: ``once an influencer - always an influencer''. In this study, we reformulate the transductive influencer detection into the inductive influencer \textit{prediction} problem for imbalanced customer networks to allow for re-influencing within the existing network nodes and generalizing towards nodes not seen by the model. We also extend the analysis by incorporating a more sophisticated GNN encoder, i.e., Graph Isomorphism Network, employing focal loss, and performing a profit-driven model evaluation, which has not been previously explored for influencer prediction.

\section{Methodology}
\label{methodology}
\subsection{Problem definition}

Following the pipeline developed by \cite{zhu2022learnable}, we define influencer prediction as a supervised node-level learning on a dynamic heterogeneous undirected network following the stacked DTDG approach. We consider a dynamic network as an ordered sequence of \textit{T} network snapshots $G = \langle G_1, G_2, ... , G_T \rangle$,  $G_t = (V_t,E_t)$ where $V_t$ is the set of nodes at time $t$, $|V_t| = n$, and $E_t \subseteq V_t \times V_t$ is the set of edges at time $t$, representing the connections between nodes, $t \in \{1, ... , T\}$. Each node is labelled with a vector $\mathbf{l}_v^t$ at time $t$ with values in $\{0, 1\}$, an influencer is labelled as 1, and vice versa.  Each node $v \in V_t$ is attributed with node features $\mathbf{x}_v^t$ at time $t$. Each edge $(u,v) \in E_t$ between nodes $u,v \in V_t$  is characterized by edge features $\mathbf{e}_{u,v}^t$ at time $t$ with values in $\{0, 1\}$. For the static problem formulation (i.e., GNN specification), $t$ is omitted. The task is node classification, i.e., predicting a label $\mathbf{l}_v^{T+k}$ for a node $v \in V_{T+k}$ at time $T+k$.

\subsection{GNN-RNN configurations}
\label{GNN_RNN}

Introduced by \cite{veličković2018graph}, GATs consist of layers in which nodes use attention mechanism to emphasize connections with neighbors. Stacking these layers gives implicit weights to neighboring nodes without requiring full knowledge of the graph. \cite{brody2022how} proposed a more expressive version, GATv2, which employs dynamic graph attention. In GATv2, the attention ranking depends on the query node, allowing all nodes to attend to each other. In a single layer of GATv2, node features as well as edge features (if any), are taken as input, and a set of node embeddings is produced as an output. To accomplish this, each node attends to its one-hop neighbors, including itself, by calculating attention coefficients $\alpha_{i,j}$ and the final node embeddings  $\mathbf{x}^{\prime}_i$ as follows:

\small
\begin{equation}
\alpha_{i,j} =
\frac{
\exp\left(\mathbf{a}^{\top}\mathrm{LeakyReLU}\left(\mathbf{\Theta}
[\mathbf{x}_i \, \Vert \, \mathbf{x}_j \, \Vert \, \mathbf{e}_{i,j}]
\right)\right)}
{\sum_{k \in \mathcal{N}(i) \cup \{ i \}}
\exp\left(\mathbf{a}^{\top}\mathrm{LeakyReLU}\left(\mathbf{\Theta}
[\mathbf{x}_i \, \Vert \, \mathbf{x}_k \, \Vert \, \mathbf{e}_{i,k}]
\right)\right)}
\end{equation}
\normalsize

\small
\begin{equation}
\mathbf{x}^{\prime}_i = \concat_{m=1}^K \sigma ( \alpha_{i,i}^{m}\mathbf{\Theta}^{m}\mathbf{x}_{i} +
\sum_{j \in \mathcal{N}(i)} \alpha_{i,j}^{m}\mathbf{\Theta}^{m}\mathbf{x}_{j})
\end{equation}
\normalsize where $\mathcal{N}(i)$ is a set of nodes adjacent to $i$, $\mathbf{\Theta}$ is a weight matrix, $\alpha_{i,i}^{m}$ is an attention coefficient computed by the \textit{m}-th attention mechanism, $\mathbf{a}^{\top}$ is a weight vector of the attention mechanism network, $\mathbin\Vert$ is concatenation, $\mathbf{x}_i$ are the features of a node $i$ and $\mathbf{e}_{i,j}$ are the features of an edge between nodes $i$ and $j$.

Recent research on GNNs has focused on maximizing the representational power of learned embeddings. GINs, introduced by \cite{xu2018powerful}, have been shown to be as powerful as the Weisfeiler-Lehman graph isomorphism test (WL test) \citep{leman1968reduction}. GINs are thus considered to be among the most powerful GNNs in terms of discriminative and representational power \citep{xu2018powerful}. Graph isomorphism is closely related to graph learning: isomorphic graphs should be mapped to the same representation, and vice versa. To achieve the same power as the WL test, a GNN's neighbor aggregation must be injective. GINs employ a Multi-Layer Perceptron (MLP) to model and learn these injective functions \citep{xu2018powerful}. We use the modified GIN operator \citep{hu2020strategies} which can incorporate edge features into the aggregation procedure, contrary to the original GIN operator \citep{xu2018powerful} that can only deal with node features, as follows:

\begin{equation}
\mathbf{x}^{\prime}_i = h_{\mathbf{\Theta}} ( (1 + \epsilon) \cdot
\mathbf{x}_i + \sum_{j \in \mathcal{N}(i)} \mathrm{ReLU}
( \mathbf{x}_j + \mathbf{e}_{j,i} ) )
\end{equation} where $h_{\mathbf{\Theta}}$ is an MLP, $\mathbf{x}_i$ is a feature vector of node $i$, $\mathbf{x}_j$ is a feature vector of node $j$ from a set of nodes $\mathcal{N}(i)$ adjacent to $i$, $\mathbf{e}_{j,i}$ is an edge features vector. $\epsilon$ determines the importance of the target node compared to its neighbors and is set to 0 following the findings of \cite{xu2018powerful} where learning $\epsilon$ yielded no gain in fitting training data.

While the aforementioned GNNs proved effective, they are static and cannot capture changes that occur over time. To address this, we incorporate RNNs into the decoder component of our framework, specifically the Long Short-Term Memory (LSTM) and Gated Recurrent Units (GRUs) models \citep{hochreiter1997long,cho2014learning}. The LSTM model aims to learn long-term dependencies with an LSTM cell consisting of several gates, including the input gate $i_t$, the forget gate $f_t$, the cell gate $g_t$, and the output gate $o_t$ \citep{hochreiter1997long}. The forget gate is used to decide which information to keep and which to discard, while the input and cell gates use the input and previous hidden state to update the cell state. The cell state $c_t$ can then be updated with the outputs of the input, cell, and forget gates as follows:
\begin{align}
\label{eq:LSTM}
i_t &= \sigma(W_{ii} \mathbf{x}^{\prime t} + b_{ii} + W_{hi} h_{t-1} + b_{hi}) \\
f_t &= \sigma(W_{if} \mathbf{x}^{\prime t} + b_{if} + W_{hf} h_{t-1} + b_{hf}) \notag \\
o_t &= \sigma(W_{io} \mathbf{x}^{\prime t} + b_{io} + W_{ho} h_{t-1} + b_{ho}) \notag \\
c_t &= f_t \odot c_{t-1} + i_t \odot \tanh (W_{ig} \mathbf{x}^{\prime t} + b_{ig} + W_{hg}h_{t-1} + b_{hg}) \notag \\
h_t &= o_t \odot \tanh(c_t) \notag 
\end{align} where $\mathbf{x}^{\prime t}$ is the output embeddings of a GNN model at time $t$, $h_{t-1}$ is the hidden state at time $t-1$, $n_t$ is the new gate used to update a hidden state using a reset gate, $W$ are learnable weights, $b$ are bias variables, $\sigma$ is the sigmoid function, $\tanh$ is the hyperbolic tangent function, $\odot$ is the Hadamard product.  

Finally, the output gate, together with the cell state, is used to update the hidden state received from the previous timestamp $t-1$, as can be seen from \eqref{eq:LSTM}. These gates allow the cell state, which can be thought of as the long-term memory of the network, to be modified at each timestamp $t$. By leveraging these operations, the LSTM can effectively capture and learn from long-term patterns.

The GRU model, which is a simpler version of the LSTM, consists of two primary gates: the reset gate $r_t$ and the update gate $z_t$. The update gate operates similarly to the input gate in LSTM models, and the reset gate functions in a manner comparable to the forget gate. Unlike the LSTM, the GRU does not have a cell state and stores long-term memory directly in the hidden states. Each hidden state is updated with the previous hidden state values and the new gate value obtained from the reset gate output as follows:
\begin{align}
\label{eq:GRU}
& r_t = \sigma(W_{ir} \mathbf{x}^{\prime t} + b_{ir} + W_{hr} h_{t-1} + b_{hr} \\
& z_t = \sigma(W_{iz} \mathbf{x}^{\prime t} + b_{iz} + W_{hz} h_{t-1} + b_{hz}) \notag \\
& n_t = \tanh(W_{in} \mathbf{x}^{\prime t} + b_{in} + r_t \odot (W_{hn} h_{t-1} + b_{hn})) \notag \\
& h_t = (1 - z_t) \odot n_t + z_t \odot h_{t-1} \notag
\end{align} where $\mathbf{x}^{\prime t}$ is the output embeddings of a GNN model at time $t$, $h_{t-1}$ is the hidden state at time $t-1$, $n_t$ is the new gate used to update a hidden state using a reset gate, $W$ are learnable weights, $b$ are bias variables, $\sigma$ is the sigmoid function, $\tanh$ is the hyperbolic tangent function, $\odot$ is the Hadamard product. 

As can be seen from \eqref{eq:LSTM} and \eqref{eq:GRU}, the input to both the LSTM and GRU models are the output embeddings $\mathbf{x}^{\prime t}_i$ of a GNN encoder produced at time $t$ for each node $i \in V_t$, with the subscript $i$ omitted for simplicity.

\begin{figure*} 
    \centering
  \subfloat[GNN-RNN framework\label{GNN-RNN}]{%
       \includegraphics[width=0.3\textwidth]{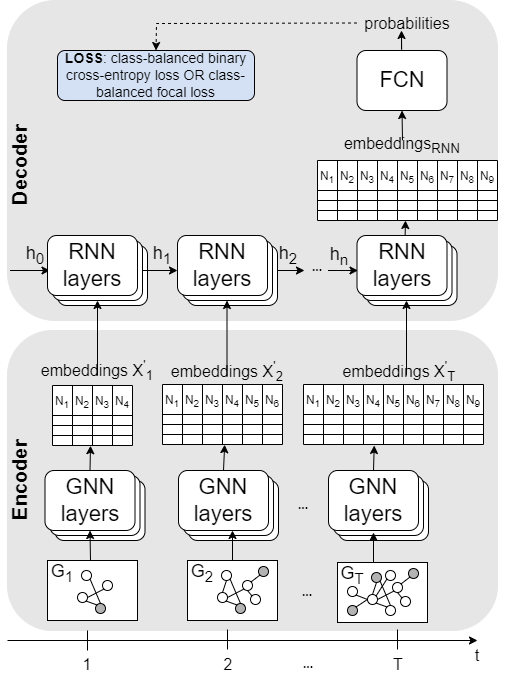}}
       \hspace{0.5cm}
  \subfloat[Features(+PR)-RNN framework\label{Features-RNN}]{%
        \includegraphics[width=0.3\textwidth]{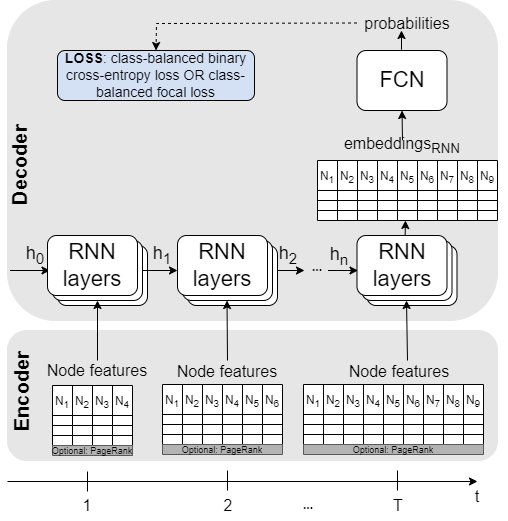}}
        \hspace{0.5cm}
  \subfloat[Static GNN framework\label{Static-GNN}]{%
        \includegraphics[width=0.3\textwidth]{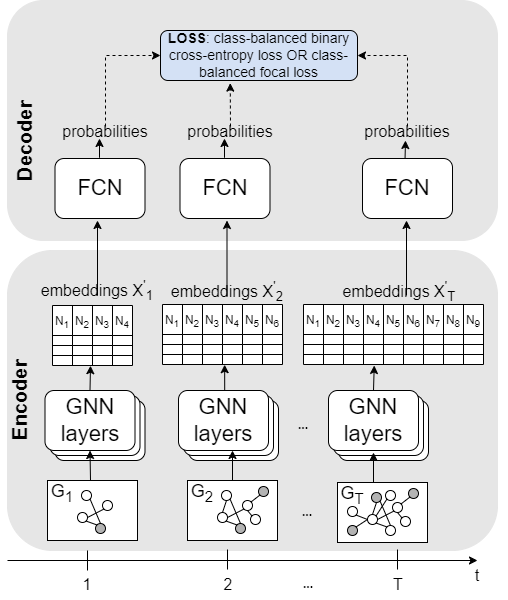}}
  \caption{Model architectures. (a) For each time point, the model takes the network as input, i.e., network connectivity, node \& edge features. This input is first passed through GNN layer(s), which generate network embeddings in Euclidean space. These embeddings, together with the hidden state from the previous step (initially a tensor of ones), are then fed into RNN layer(s). This process is repeated for each time point. At the last time point, the output of the RNN model is fed into a fully connected layer to produce the final probabilities of a node being an influencer. The weights of these layers are learned by optimizing a loss function, using either class-balanced binary cross-entropy loss or class-balanced focal loss (see Section \ref{loss_functions}). (b) For each time point, the model takes node features as input (with the PageRank feature added as well in the corresponding configurations). These features, along with the hidden state from the previous step (initially a tensor of ones), are then fed into the RNN layer(s). This process is repeated for each time point. At the final time point, the output of the RNN model is fed into a fully connected layer to produce the final probabilities of a node being an influencer. The weights of these layers are learned by optimizing a loss function, using either class-balanced binary cross-entropy loss or class-balanced focal loss (see Section \ref{loss_functions}). (c) For each time point, the model takes the network as input, i.e., network connectivity, node \& edge features. This input is passed through GNN layer(s) that generate network embeddings in Euclidean space. The embeddings are then passed to a Fully Connected Layer to produce the final probabilities of a node being an influencer. The weights of these layers are learned by optimizing a loss function, using either class-balanced binary cross-entropy loss or class-balanced focal loss (see Section \ref{loss_functions}). There is no temporal component in this configuration, and time points are used in a batch manner.}
\end{figure*}

To obtain final predictions, we require an additional Fully Connected Network (FCN) on top of the GNN-RNN configuration. The FCN consists of two linear layers followed by ReLU and sigmoid activation functions, respectively. It takes as input the node embeddings, i.e., the hidden states of the last output layer of the RNN model, and outputs the probability of a node being an influencer. The architecture is shown in Figure \ref{GNN-RNN}. 

\subsection{Baseline configurations}
\paragraph{Features(+PR)-RNN} 
Following the strategy of \cite{oskarsdottir2017social}, we can enrich non-relational classifiers with network features. One such network feature that can summarize the importance of a node is PageRank (PR) \citep{brin1998anatomy}. Originally developed to measure the importance of web pages, PR can be used to measure the importance of nodes based on the number of incoming edges and their importance. The PageRank measure and its variants have been shown to boost predictive models \citep{oskarsdottir2022social}. The non-personalized PR values can be calculated for each node as 
$PR(u) = (1-d) + d \sum_{v \in B_u} \frac{{PR(v)}}{{C(v)}}$ where $B_u$ is a set of nodes linked to $u$, $d$ is a damping factor set to 0.85, $C(v)$ is an out-degree of a node $v$. 

Since the PR value represents the relative importance of a node within one component, we calculate its value separately within each of the components and scale the value with the size of the component. Since the network is attributed, PR values can be seen as an additional feature of the node, and this enriched feature set can be seen as an encoder part and can be subsequently used for dynamic node classification with RNNs. Static node features together with PR are used as input to the RNN models, i.e., LSTM and GRU models. The aforementioned architecture is shown in Figure \ref{Features-RNN}. 

\paragraph{Static GNN}
\label{static_GNN}

This configuration consists of a GNN encoder similar to the GNN-RNN configuration described in Section \ref{GNN_RNN}, while its decoder contains only an FCN network that is fed with the embeddings produced by the GNN model. For the GNN encoder, we use GAT and GIN models described in Section \ref{GNN_RNN}. The aforementioned architecture is shown in Figure \ref{Static-GNN}.

\section{Experimental setup}
\label{experiments}
\subsection{Networks}
\label{network}

The data used in this study is sourced from a Super-App company operating in Latin America that offers both a delivery app and credit card services. The data covers a period of nine months, during which aggregated information on delivery app usage was used to create monthly snapshots of the network, while credit card usage data was used to construct node features (see Appendix \ref{appen:node_features}, Table \ref{node_features}). Numerical features are normalized using min-max normalization to stabilize learning and speed up convergence. Categorical features are one-hot encoded. To account for new nodes appearing in the network over time, we add artificial node features for the future nodes (not used in the backpropagation process during training) with features that are zeroed out. The network only includes credit card customers who use the delivery app, with three types of connections based on the app usage, as described in Table \ref{Edge_types}. To account for these types of connections, edge features are one-hot encoded with information about the connection type(s), allowing multiple connection types to be represented per edge. The network is dynamic, with new nodes and edges appearing over time, while existing nodes and connections persist, leading to its continuous growth. 

\begin{table}
\fontsize{8}{10}\selectfont 
\caption{Edge types}
\label{Edge_types}
\centering
\begin{tabular}{p{2cm}p{10cm}}
\hline
\textbf{Type} & \textbf{Description}\\
\hline
Credit card & There exists an edge between customers who used the same credit card in the delivery app.\\
Geohash & There exists an edge between customers who ordered more than 4 times in the close geographical proximity (Geohash 7–152.9m x 152.4m).\\
Contact & There exists an edge between customers if one of them has the other one in their phone contacts. \\
\hline
\end{tabular}
\end{table}

Our analysis is based on data from three different cities, each represented by a separate network (network characteristics are provided in Appendix \ref{appen:networks}, Figure \ref{fig:networs_char}). The cities are diverse in terms of population size and income, representing both middle-income and high-income cities, with Purchase Power Parity-adjusted Gross Domestic Products (PPP GDP) per capita ranging from \$10,000 to \$35,000\footnote{For privacy reasons, we cannot disclose the names of the cities.}. The networks vary in size, with City 1 having the smallest network and City 3 having the largest. Over time, all three networks have grown in terms of both the number of nodes and the number of edges. However, the data is heavily imbalanced as the proportion of influencers among the nodes is relatively low, ranging from about  1.5\% in the last month to about 5\% in the first month. To address this issue, we use the synthetic node generation method of GraphSMOTE, which performs oversampling on top of the embedding space constructed by the feature extractor \citep{zhao2021graphsmote}. The original GraphSMOTE constructs an embedding space to encode the similarity between the nodes, and uses this space to generate new samples, while also generating edges to model the relationship information for these new samples \citep{zhao2021graphsmote}. We use a part of GraphSMOTE that interpolates the generated embeddings of the minority class to create new synthetic embeddings of that class. In particular, we oversample the final embeddings generated by the RNN model for the GNN-RNN model configurations or by the GNN encoder in static GNN configurations with an oversampling scale of 0.5, as scales below 0.8 typically lead to better GNN performance \citep{zhao2021graphsmote}. In particular, we have included a positive oversampling ratio as a hyperparameter specification, while also allowing for a non-oversampling strategy with a ratio of 0.

\subsection{Influencer definition}
\begin{figure*}
\centering
\includegraphics[width=12cm]{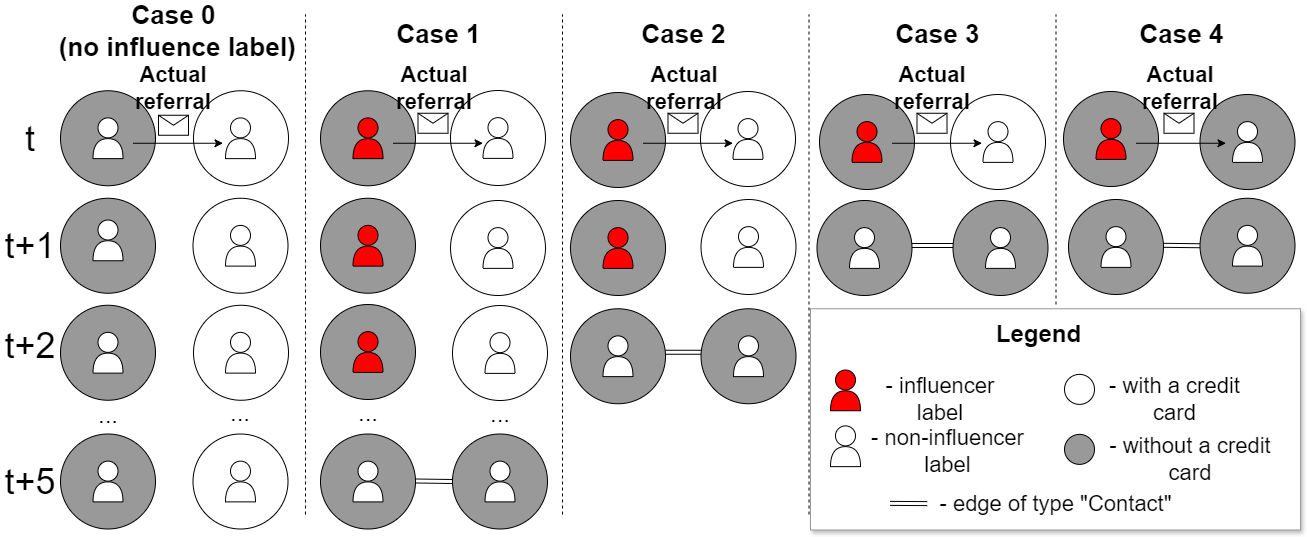}
\caption{Labelling intuition}
\label{Labelling}
\end{figure*}

An influencer is defined as a super-app user with a credit card who successfully refers another super-app user without a credit card, resulting in the referred person obtaining a credit card. In particular, we use the following intuition for node labeling. Node labels are determined based on referral information from credit card applications. A referrer is always a node in the network, while the referred person may or may not be in the network, depending on when they obtain a credit card. A node is labeled as an influencer at the time of the actual referral and remains labeled as such until the referred person obtains a credit card, which is the point at which the company has access to their usage data, up to a maximum of six months. At that point, the influencer label is changed to a non-influencer label, unless the customer has made other referrals in the past six months, as obtaining a credit card is no longer influenced solely by a referral, but by other factors as well, and the information asymmetry between referred and non-referred customers vanishes over time \citep{van2018customer}. To maintain the connectivity expressed by a referral, an edge of type "Contact" is created between the referrer and the referred person by adding the referred person as a node with an edge to the referrer. Figure \ref{Labelling} illustrates the labeling process. We do not distinguish between the influencing strength (e.g., whether one node is considered more influential than another one) as we are dealing with binary classification.

\subsection{Loss functions}
\label{loss_functions}
As part of our model optimization, we explore two loss function configurations: weighted binary cross-entropy and class-balanced focal loss. A traditional binary cross-entropy (BCE) loss is not suitable for handling class imbalance, since it treats both classes equally and prioritizes overall accuracy rather than class-specific performance. Therefore, a custom binary cross-entropy loss function is needed to handle class imbalance and weight minority examples accordingly to balance the trade-off between recall and precision, as follows:

\small
\begin{align}
\label{bce}
\ell_c(x,y) &= \sum_{i=1}^{n} \left(-w_{n,c} [ \, p_c y_{n,c} \cdot \log (\sigma(x_{n,c})) \right. \nonumber \\
&\quad\quad\quad\quad\quad\quad+ \left. (1-y_{n,c}) \cdot \log (1-\sigma(x_{n,c})) ] \right)
\end{align}
\normalsize where $c=1$ for single-label binary classification, $n$ is the number of samples, $p_c = \frac{n_{neg}}{n_{pos}}$ is the weight of the positive class, $n_{neg}$ ($n_{pos}$) is the number of samples in the negative (positive) class, $\sigma$ is a sigmoid function. 

To go beyond simple inverse sample weighting, \cite{cui2019class} proposed a class-balanced focal loss, which addresses the challenge of class imbalance in a more sophisticated way by considering both the effective number of samples and the relative loss for well-classified samples \citep{lin2017focal,cui2019class}, as follows:

\small
\begin{equation}
\label{focal}
\ell_c(x,y) = \sum_{i=1}^{n} \left( - \frac{1-\beta}{1-\beta^{n_c}} \sum_{i=1}^{C} (1-\sigma(x_{n,c}))^\gamma \log (\sigma(x_{n,c})) \right)
\end{equation} 
\normalsize where $C$ is the total number of classes, $n_c$ is the number of samples in the ground-truth class $c$, $\beta \in [ \,0,1)$ is a hyperparameter from the effective number of samples for class term, $\gamma$ is a focusing hyperparameter. 

By adjusting a class-balanced term between no reweighting and reweighting by inverse class frequency, the class-balanced focal loss is able to handle class imbalance more effectively \citep{cui2019class}.

\subsection{Hyperparameter tuning}
\label{hyperparameters}
The best model configuration is found by an exhaustive grid search over the hyperparameter space (Table \ref{Hyperparameters}). We select the best model based on the average AUPRC value calculated on seen and unseen nodes from the validation set as well as based on stability of the training process, i.e., we ensure that the model converges. In addition, the number of heads in GAT is fixed at 2 to keep the computational complexity manageable. We also set the focal loss hyperparameters (see Section \ref{loss_functions}) $\gamma$ and $\beta$ to 2 and 0.999, respectively, as these values have been found to perform best \citep{lin2017focal,cui2019class}. For both the GNN-RNN and Static GNN models, we applied two regularization techniques that are commonly used for GNNs to improve their performance and generalization capabilities: layer normalization and a 50\% dropout rate. 

 \begin{table}[h]
\fontsize{8}{10}\selectfont 
\caption{Hyperparameters tuning\tablefootnote{For City 3, GAT-RNN models, the following configurations are not feasible due to computing limitations: all models with GNN layers = 4; configurations with GNN hidden dimensions = RNN hidden dimensions = GNN embeddings = 512}. The models are trained on NVidia A100 GPU for 300 epochs with a learning rate of 0.0001 using the ADAM optimizer \citep{DBLP:journals/corr/KingmaB14}. The models are implemented using Pytorch Geometric \citep{Fey/Lenssen/2019}.}
\label{Hyperparameters}
\centering
\begin{tabular}{p{1.5in}p{1in}p{1.3in}}
\toprule
Hyperparameter & Range & Model \\
\midrule
GNN hidden dimensions, GNN embeddings & 256, 512 & GNN-RNN, Static GNN \\
GNN layers & 2, 4 & GNN-RNN, Static GNN \\
RNN hidden dimensions & 256, 512 & GNN-RNN, Features(+PR)-RNN \\
RNN layers & 1, 2 & Features(+PR)-RNN \\
SMOTE sample rate & 0, 0.5 & All models \\
Loss function & BCE Loss, Focal Loss & All models \\
\bottomrule
\end{tabular}
\end{table}

\subsection{GNN-RNN models training}
\label{GNN_RNN_training}
\begin{figure*} 
    \centering
  \subfloat[GNN-RNN and Features (+PR)-RNN models\label{setup}]{%
       \includegraphics[width=0.54\textwidth]{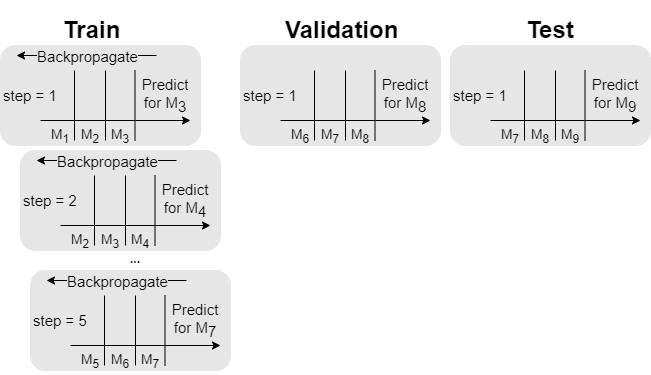}}
       \hspace{0.5cm}
  \subfloat[Static GNN models\label{Static_GNN_setup}]{%
        \includegraphics[width=0.41\textwidth]{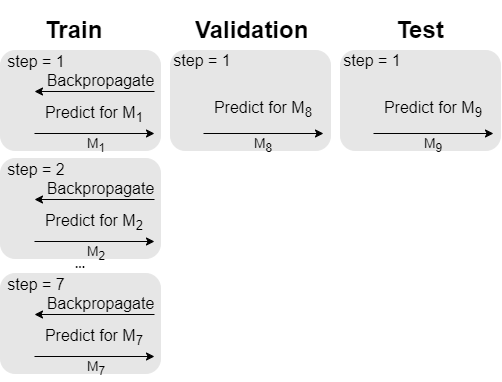}}
  \caption{Train/validation/test setup}
\end{figure*}

To evaluate model performance and ensure that it can generalize, we use a rolling window strategy with a window size of three months and a shift of one month to split the data into train, validation, and test subsets according to the out-of-time validation and testing strategy (Figure \ref{setup}). 

The train data spans seven months, resulting in five three-month windows. Within each window, the trained model is used to make predictions for the last month. We use a network snapshot of each month as input to the GNN encoder, which produces embeddings that are subsequently fed to the RNN decoder, with the input hidden states initialized as a tensor of ones for the first time window. For all subsequent time windows, the hidden state produced by the first month of the previous time window is used as the input hidden state. The output produced by the RNN decoder at time $t$ is used as the input hidden state at time $t+1$. At the final timestamp of the window, the hidden states produced by the RNN decoder are fed into the FCN decoder to generate final probabilities, and backpropagation occurs during training. 

Both validation and test data consist of one three-month window. The trained models are applied to a validation set that is used to measure model generalization capability by checking its convergence during training by inspecting the validation loss. Additionally, a validation set is used to select the best hyperparameter configuration (see Section \ref{hyperparameters}). A test set is used for the final model comparison (with the best hyperparameter setting chosen on a validation set).

\subsection{Baseline models training}

The process of splitting the data into train, validation, and test subsets and the training methodology for the Features(+PR)-RNN models are similar to those of the GNN-RNN models, with the only difference being the encoder part. In this case, instead of using GNN embeddings, the encoder consists only of node features (including the PageRank feature for PR configurations), which are fed directly into the RNN decoder. The rest of the training process follows the steps outlined in Section \ref{GNN_RNN_training} (Figure \ref{setup}).

Unlike the previous models with the RNN component, static GNN models do not take dynamics into account. The training data for the static model spans seven months, with predictions made for each month and the training performance averaged over those months. The validation and test sets consist of data from a single month each, with predictions made for their respective months (Figure \ref{Static_GNN_setup}).

\subsection{Performance metrics}
\label{performance_metrics}

We are solving the influencer prediction problem in a profit-driven manner, and therefore the final model threshold decision is made based on profit rather than predictive performance at that particular threshold. Therefore, we use threshold-independent metrics.

The area under the receiver operating characteristic (ROC) curve (AUC), which plots the true positive rate (TPR) against the false positive rate (FPR) for different classification thresholds, is often calculated to compare model performance, with a range of values from 0 to 1 and a baseline of 0.5. In the presence of class imbalance, AUC may not be a reliable performance indicator as it is necessary to consider the skewed class distribution. The Precision-Recall (PR) curve is a better alternative for imbalanced data as it summarizes performance over different thresholds with precision and recall on its axes \citep{davis2006relationship}. To facilitate model comparison, the area under the PR curve (AUPRC) can be calculated with the baseline corresponding to the fraction of positives. While the AUPRC metric is our primary criterion for evaluating performance, we also report the AUC metric. 

For both AUC and AUPRC, we report the values separately for seen and unseen nodes. The term ``seen nodes'' refers to nodes that the model encountered during backpropagation in training, whereas ``unseen nodes'' refers to nodes that only appear in the network later. Hence, ``unseen nodes'' for the validation data are the nodes that appeared in the last month of the validation window, while the ``unseen nodes'' for the test data include new nodes that appeared in the last month of the test window as well as in the last month of the validation window, as all of them were not part of the training. Model performance on seen nodes shows the models' applicability for transductive learning which focuses on making predictions specifically for the given training instances without generalizing to unseen instances, while performance on unseen nodes highlights the ability to generalize from observed data to make predictions on unseen instances, i.e., inductive learning.

We also report training time and carbon footprint estimate measured in equivalent grams of \ce{CO_2} (\textit{gCO2eq}), as we also want to evaluate the tradeoff between energy efficiency and prediction performance \citep{anthony2020carbontracker}.

\section{Results}
\label{results}

Table \ref{Models_performance} displays the values of the performance metrics. The results for City 1 demonstrate that GIN-RNN models  perform similarly to dynamic non-GNN models in terms of AUC for both seen and unseen nodes, with GIN-GRU being the best performing model for seen nodes. The AUPRC values for seen nodes are also comparable across these model types, with GAT-LSTM achieving the highest value. Notably, the Static GIN model delivers exceptional performance in terms of AUC on unseen nodes. Nevertheless, the highest AUPRC values for unseen nodes are obtained by the GIN-LSTM and GIN-GRU models. The GIN-RNN models achieve comparable time performance to the baseline models, while the GAT-RNN models have the longest training time and produce the highest amount of $CO_2$ emissions due to an expensive attention computation. 

\begin{table}[h!]
    \centering
      \caption{Models performance on test data. Best model configurations can be found in Tables \ref{tab:model_configurations_city1}-\ref{tab:model_configurations_city3}}
\fontsize{6}{8}\selectfont 
  \begin{tabular}{p{3cm}p{1.3cm}p{1.3cm}p{1.3cm}p{1.3cm}p{1.1cm}p{1.1cm}}
    \toprule                 
    Model & AUC seen nodes & AUC unseen nodes & AUPRC seen nodes & AUPRC unseen nodes & Total time, s. & $CO_2$, g. \\
    \midrule
    \multicolumn{7}{c}{City 1} \\
    \midrule
    GIN-LSTM  & 0.75$\pm$0.02 & 0.67$\pm$0.06 & 0.08$\pm$0.02 &  \textbf{0.11$\pm$0.06} & 57 & 5.21 \\
    GIN-GRU  & \textbf{0.79$\pm$0.02} & 0.63$\pm$0.06 & 0.09$\pm$0.02 &  \textbf{0.10$\pm$0.06} & 48 & 3.68 \\
    GAT-LSTM  & 0.70$\pm$0.03 & 0.60$\pm$0.07 & \textbf{0.12$\pm$0.03} &  0.07$\pm$0.04 & 188 & 16.29 \\
    GAT-GRU  & 0.65$\pm$0.03 & 0.55$\pm$0.06 & 0.08$\pm$0.02 &  0.05$\pm$0.04 & 183 & 17.12 \\
    Features(+PR)-LSTM & 0.77$\pm$0.02 & 0.68$\pm$0.06 & 0.10$\pm$0.03 &  0.08$\pm$0.05 & 45 & 3.63 \\
    Features(+PR)-GRU & 0.77$\pm$0.02 & 0.67$\pm$0.06 & 0.09$\pm$0.02 &  0.09$\pm$0.05 & 41 & 3.01 \\
   Features-LSTM & 0.75$\pm$0.02 & 0.65$\pm$0.07 & 0.09$\pm$0.02 &  0.08$\pm$0.05 & 90 & 5.88 \\
   Features-GRU & 0.77$\pm$0.02 & 0.68$\pm$0.06 & 0.09$\pm$0.02 &  0.09$\pm$0.05 & 43 & 2.75 \\
   Static GIN & 0.69$\pm$0.03 & \textbf{0.72$\pm$0.06} & 0.05$\pm$0.01 &  0.09$\pm$0.05 & 36 & 2.21 \\
   Static GAT & 0.64$\pm$0.03 & 0.54$\pm$0.06 & 0.03$\pm$0.004 &  0.03$\pm$0.02 & 94 & 9.27 \\
   \midrule
   \multicolumn{7}{c}{City 2} \\
   \midrule
    GIN-LSTM  & \textbf{0.71$\pm$0.02} & 0.78$\pm$0.04 & \textbf{0.08$\pm$0.02 }& \textbf{0.19$\pm$0.08} & 54 & 3.90 \\
   GIN-GRU  &  0.67$\pm$0.02 & \textbf{0.80$\pm$0.04} & \textbf{0.07$\pm$0.02} & \textbf{0.22$\pm$0.08} & 61 & 4.31 \\
   GAT-LSTM  &  0.68$\pm$0.03 & 0.74$\pm$0.05 & 0.06$\pm$0.01 & 0.16$\pm$0.07 & 104 & 9.21 \\
    GAT-GRU  & 0.66$\pm$0.03 & 0.73$\pm$0.06 & 0.06$\pm$0.02 & 0.18$\pm$0.08 & 202 & 18.82 \\
     Features(+PR)-LSTM & 0.70$\pm$0.02 & 0.64$\pm$0.07 & 0.04$\pm$0.01 & 0.07$\pm$0.04 & 47 & 3.54 \\
    Features(+PR)-GRU & \textbf{0.71$\pm$0.02} & 0.65$\pm$0.07 & 0.06$\pm$0.02 & 0.07$\pm$0.03 & 38 & 2.33 \\
   Features-LSTM & 0.70$\pm$0.02 & 0.64$\pm$0.07 & 0.05$\pm$0.01 & 0.07$\pm$0.03 & 48 & 3.82 \\
    Features-GRU & \textbf{0.71$\pm$0.02} & 0.66$\pm$0.07 & 0.05$\pm$0.01 & 0.08$\pm$0.04 & 99 & 5.86 \\
   Static GIN & 0.66$\pm$0.03 & \textbf{0.83$\pm$0.03} & 0.05$\pm$0.01 & 0.13$\pm$0.06 & 37 & 2.20 \\
   Static GAT & 0.60$\pm$0.03 & 0.70$\pm$0.05 & 0.03$\pm$0.003 & 0.07$\pm$0.04 & 60 & 4.83 \\
   \midrule
   \multicolumn{7}{c}{City 3} \\
   \midrule
   GIN-LSTM  & 0.73$\pm$0.01 & \textbf{0.78$\pm$0.02} & 0.08$\pm$0.01 & \textbf{0.13$\pm$0.03} & 517 & 40.61 \\
   GIN-GRU  &  \textbf{0.75$\pm$0.01} & \textbf{0.78$\pm$0.02} & \textbf{0.09$\pm$0.01} & \textbf{0.13$\pm$0.02} & 622 & 57.51 \\
   GAT-LSTM  &  0.69$\pm$0.01 & 0.68$\pm$0.02 & \textbf{0.09$\pm$0.01} & 0.12$\pm$0.02 & 685 & 70.90 \\
   GAT-GRU  & 0.68$\pm$0.01 & 0.70$\pm$0.02 & \textbf{0.09$\pm$0.01} & \textbf{0.13$\pm$0.03} & 525 & 45.06 \\
     Features(+PR)-LSTM & 0.70$\pm$0.01 & 0.67$\pm$0.02 & 0.03$\pm$0.002 & 0.03$\pm$0.006 & 433 & 34.71 \\
   Features(+PR)-GRU & 0.71$\pm$0.01 & 0.66$\pm$0.02 & 0.04$\pm$0.003 & 0.05$\pm$0.01 & 5106 & 302.56 \\
   Features-LSTM & 0.69$\pm$0.01 & 0.66$\pm$0.02 & 0.03$\pm$0.002 & 0.03$\pm$0.005 & 2823 & 166.58 \\
    Features-GRU & 0.71$\pm$0.01 &0.67$\pm$0.02 & 0.04$\pm$0.002 & 0.04$\pm$0.01 & 397 & 25.44 \\
    Static GIN & 0.69$\pm$0.01 & 0.72$\pm$0.02 & 0.05$\pm$0.004 & 0.10$\pm$0.02 & 360 & 22.87 \\
   Static GAT & 0.63$\pm$0.01 & 0.61$\pm$0.02 & 0.04$\pm$0.003 & 0.12$\pm$0.03 & 909 & 81.86 \\
    \bottomrule
  \end{tabular}
  \label{Models_performance}
\end{table}

For the larger network of City 2, the GNN-RNN models obtain similar AUC results as the dynamic non-GNN models on seen nodes. However, the best AUC values on unseen nodes are obtained by the static GIN model, followed by the GIN-GRU model. The GIN-RNN models show the best AUPRC performance on both seen and unseen nodes, with a strong dominance on unseen nodes and an AUPRC uplift of about 0.12 from the best performing dynamic non-GNN model. Overall, the GIN-GRU model provides the best performance based on most of the metrics. In terms of time and energy performance, the pattern is the same as for City 1: complex GAT-RNN models take the longest time to train and produce maximum carbon emissions.

For the largest network of City 3, it becomes apparent that the GNN-RNN models outperform the baseline models. In particular, the GIN-RNN models achieve the highest AUC values on both seen and unseen nodes, indicating their superior performance. Additionally, in terms of AUPRC on seen and unseen nodes, all GNN-RNN models perform exceptionally well, with no clear winner among them. 

In summary, more sophisticated models, and in particular the GIN-GRU combination, perform better as the network grows, becoming dominant in the largest network of all. This is consistent with other results in deep learning, that confirm how data-hungry these models can be: in order to improve on simpler alternatives, it is necessary to have enough data diversity to achieve effective results. In terms of computational time, models with pure attention mechanisms are the most complex to train, which is consistent with their quadratic complexity. In our models, this does not translate into better performance, so modelers should carefully test other alternatives to benchmark the right model. Interestingly, the use of a balanced error measure, AUPRC, shows that more powerful models do present better results. As discussed earlier, the AUC performance can be misleading in the presence of high imbalance, so measures such as AUPRC can be more helpful.

\begin{figure*} 
    \centering
  \subfloat[Train/Val. losses\label{fig:losses}]{%
       \includegraphics[width=0.4\textwidth]{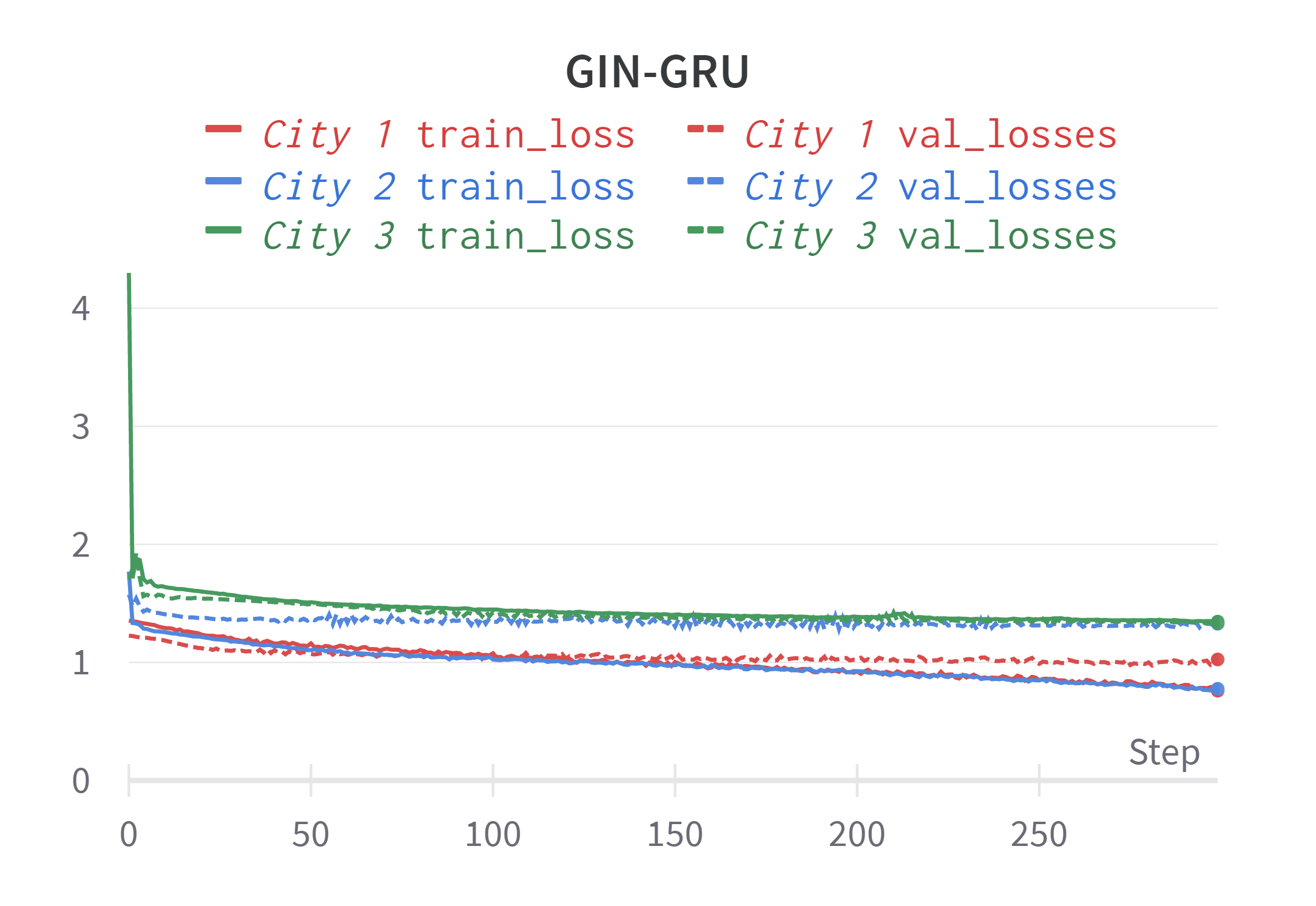}}
       \hspace{0.1cm}
  \subfloat[Validation AUC\label{fig:auc}]{%
        \includegraphics[width=0.4\textwidth]{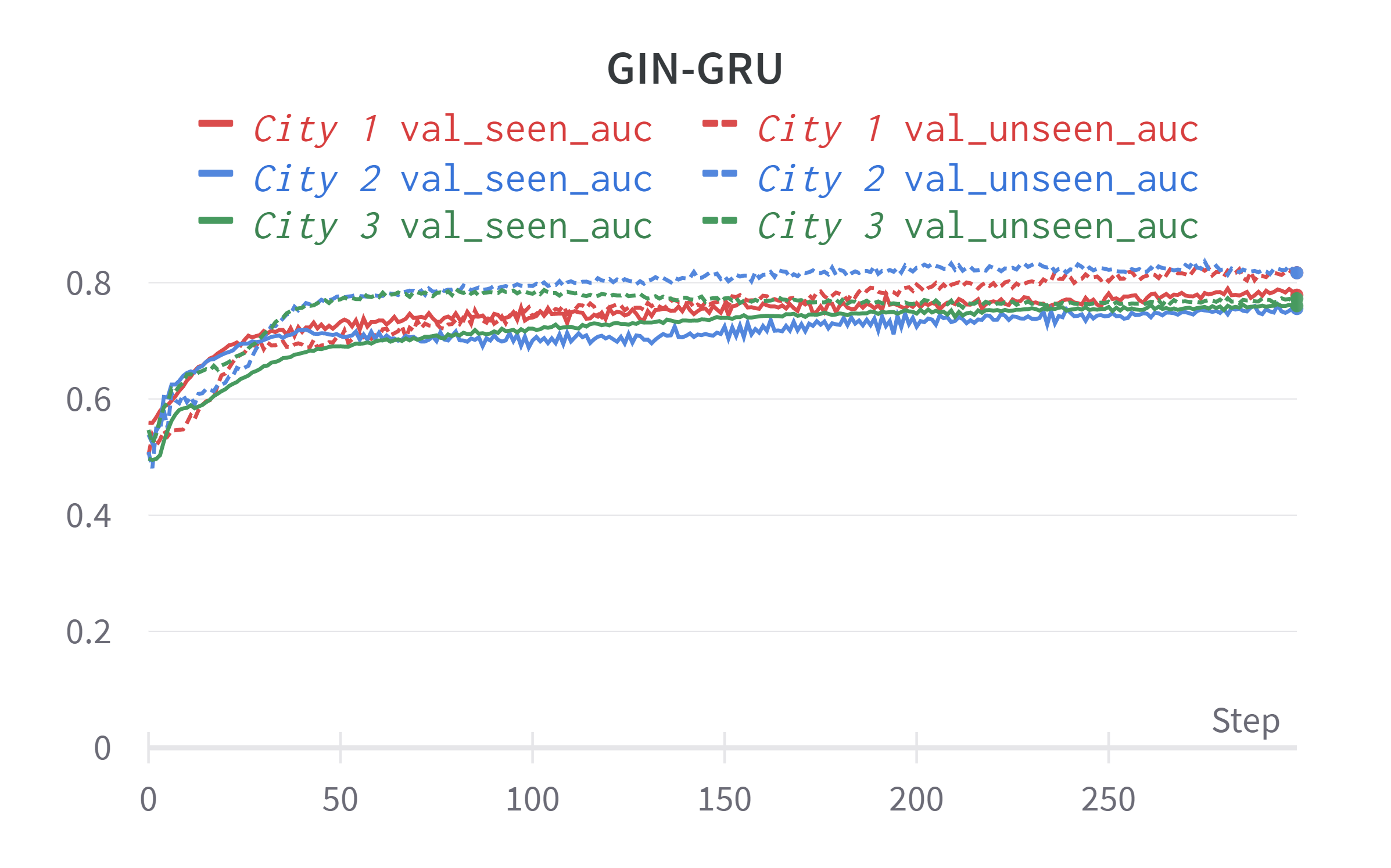}}
        \hspace{0.1cm}
\subfloat[Validation AUPRC\label{fig:auprc}]{%
        \includegraphics[width=0.4\textwidth]{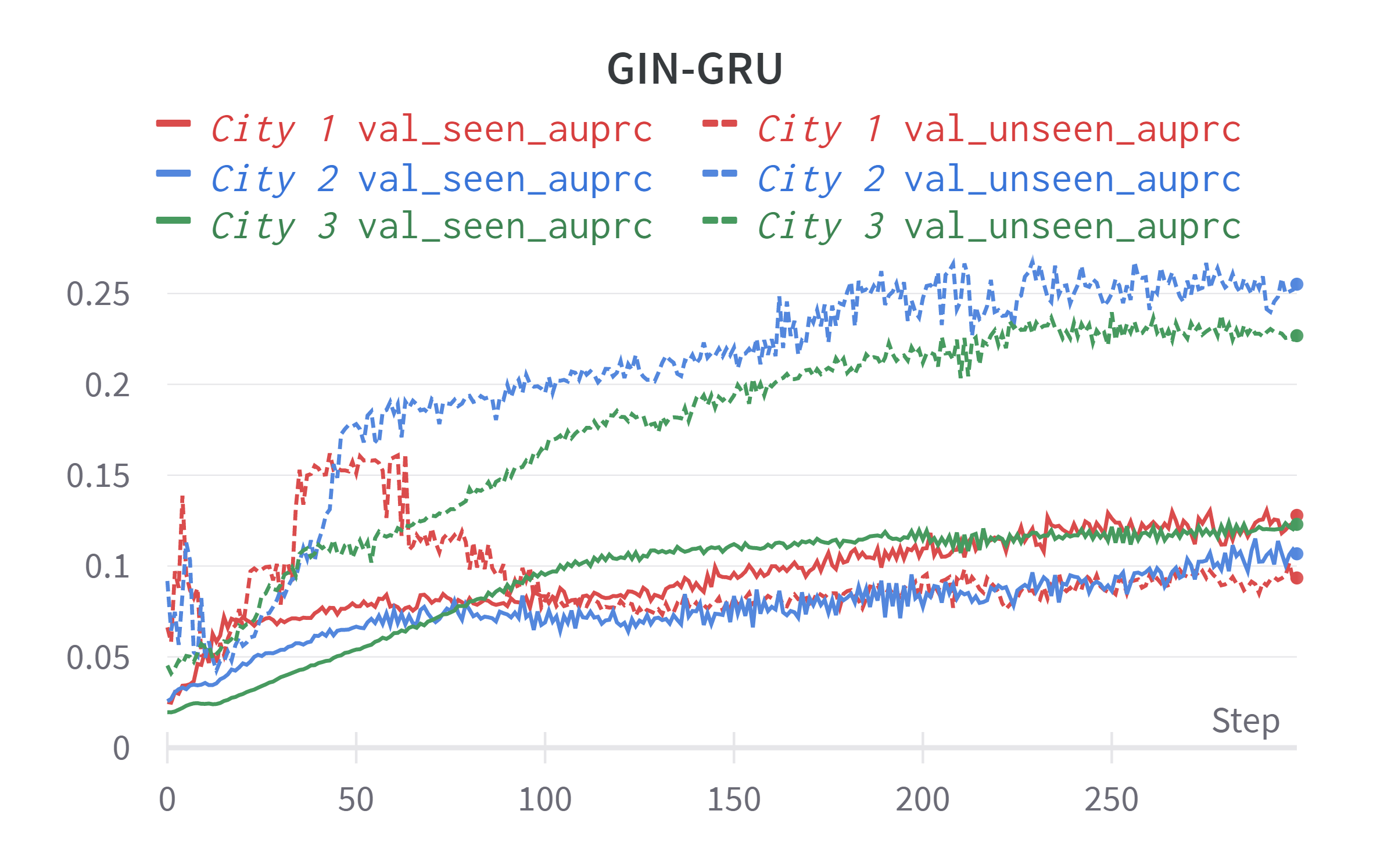}}
  \caption{Convergence plots, GIN-GRU models}
  \label{fig:convergence} 
\end{figure*}

Figures \ref{fig:losses}, \ref{fig:auc}, and \ref{fig:auprc} display train/validation losses, AUC, and AUPRC convergence plots for the consistently well-performing GIN-GRU models for all three cities. These plots demonstrate that the GIN-GRU model achieves convergence across all networks. While the validation AUC grows consistently with epochs for all cities, Figure \ref{fig:auprc} reveals that the model training for the largest city, City 3, is the most stable, with AUPRC gradually increasing with each epoch. In contrast, the validation AUPRC for City 1 is the least stable. Therefore, providing more data to the model improves both its training stability and accuracy. The convergence plots for the other models can be found in Online Appendix A. 

Appendix \ref{app:model_configurations}, Tables \ref{tab:model_configurations_city1}-\ref{tab:model_configurations_city3} present the optimal hyperparameter configurations for the GNN-RNN models and the benchmark models. Across all cities, focal loss is generally the preferred loss function for most models. The optimal number of layers in the GNN encoder shows no clear pattern for City 1, while for Cities 2 and 3, shallower networks with two layers are favored. A similar trend is observed for the RNN model: shallower networks with one layer are preferred for Cities 1 and 3, whereas no distinct pattern is evident for City 2.

There is no consistent preference for the number of hidden dimensions and embeddings in the GNN encoder, although configurations with 512 hidden dimensions are slightly favored for GNN-RNN architectures. The optimal number of hidden dimensions for the RNN component increases with data size: 256 dimensions are preferred for City 1, and 512 dimensions for City 3. Regarding the SMOTE rate, no oversampling typically results in better performance for most models. Notably, none of the GNN-RNN models benefit from oversampling, with zero rate chosen for all GNN-RNN combinations.

The analysis indicates that optimal model configurations are highly dependent on the underlying data and its specific characteristics. Although some hyperparameters, such as focal loss and the absence of oversampling, remain consistently optimal across the cities, most hyperparameters vary between the models built for different cities. Therefore, each city requires a tailored model that is trained and optimized based on its own data, as we have not been able to establish the generalizability of model configurations across cities.

\section{Profit evaluation}
\label{profit_evaluation}

To make use of the model predictions, the decision on the classification threshold must be made in order to return the final predicted labels used for a marketing campaign. When dealing with severe class imbalance, the default threshold of 0.5 used for the balanced tasks can lead to poor performance \citep{fernandez2018learning}. Therefore, we tune this threshold by applying the predict-then-optimize approach introduced by \cite{vanderschueren2022predict}: a cost-insensitive influencer detection model is used to optimize decision making, i.e., tuning a classification threshold to obtain the maximum expected profit of potential customers brought by influencers. Since profit is customer-specific, we use instance-dependent cost-sensitive thresholding \citep{vanderschueren2022predict}.

We assume that the influencer detection model is used for targeted marketing, i.e., the influencers predicted by the model are contacted with a contact cost $c$. True influencers are also rewarded at a cost of $ic$. As we know from the data provider, all the customers have been targeted during a time frame of the study. However, the assumption that this is permanent is unrealistic, so we need to model the impact of a targeted policy. Hence, we can introduce a probability $p$ of not referring new customers while having the opportunity to do so (being an influencer). Since we know that everyone has been targeted, true influencers have a probability of not referring equal to 0 ($p=0$). For non-influencers, this probability is a free parameter to be estimated ($p \neq 0$). Consequently, $(1-p)$ represents the probability that an actual influencer will still bring new customers even without being targeted. 

\begin{table*}[h]
\fontsize{8}{10}\selectfont 
  \caption{Profit. $N_u$ is the set of people referred by $u$, $c$ is a contact cost, $ic$ is a reward cost, $P_v$ is a future 6-month profit of a customer $v$ calculated as an interest earned on the outstanding balance on a credit card; p is a probability of not referring new customers while having the opportunity to do so (being an influencer).}
  \label{profit}
  \setlength{\tabcolsep}{3pt}
  \centering
  \begin{tabular}{lccc}
\hline
\multicolumn{2}{l}{\multirow{2}{*}{}}                                 & \multicolumn{2}{c}{\textbf{Actual label}} \\
\multicolumn{2}{l}{}                                                  & \textit{Influencer}    & \textit{Non-influencer}   \\ \hline
\multicolumn{1}{c}{\multirow{2}{*}{\textbf{Predicted label}}} & \textit{Influencer}     &  $P_u(TI) = \sum_{v \in N_u} \Bigl(P_v - ic \Bigl) - c$           & $P_u(TNI) = -c$             \\
\multicolumn{1}{c}{}                                 & \textit{Non-influencer} &  $P_u(NTI) = (1-p) \sum_{v \in N_u} \Bigl(P_v - ic \Bigl)$           & $P_u(NTNI) = 0$ \\ \hline      
  \end{tabular}
\end{table*}

The total profit of the campaign depends on the accuracy of the model's predictions and comprises the profit of four different categories, each constructed based on the actual and predicted influencer labels (Table \ref{profit}). The first category is the Target Influencer (TI) category, which corresponds to the True Positive predictions. The profit of this category is calculated as the future profit of potential customers $P_u$ referred by an influencer $u$, reduced by the incentive cost $ic$. In addition, a contact cost $c$ is subtracted once, as the action to contact the customer may have a cost\footnote{In an app-based contact strategy, this cost can be very marginal. In traditional marketing, this cost can escalate quickly. Users must calibrate these values based on their situation.}. The $TNI$ category represents False Positive predictions: the customers who were targeted but did not refer other customers. The profit from this category is negative because we spend money to contact them and do not get new customers in return. The $NTI$ category consists of False Negatives: true influencers who were not contacted. We expect to get profit from this category of customers with the aforementioned probability $1-p$, minus the reward cost of $ic$. The last category $NTNI$ represents True Negative predictions that bring neither profit nor cost. We search for an optimal threshold that maximizes profit on the validation data, and then use this threshold on the test data to obtain a profit estimate. Hence, we solve the optimization problem as follows:
\begin{equation}
\argminA_t P_{total}(t) = \{t \mid P_{total}(t) = \max_{t'} P(t')\}
\end{equation} where $t$ is a classification threshold, $P_{total}$ is a total profit calculated as $P_{total} = \sum_{S \in G} \sum_{u \in S} P_u, G = \{TI, TNI, NTNI, NTI\}$.

We illustrate the aforementioned framework using the best performing GIN-GRU model for City 2. The $P_{total}$ calculation has three parameters $ic$, $c$, and $p$ that influence the resulting profit. We obtained a reward parameter from the data provider and normalized it to a generic value of 100 monetary units. All values are expressed based on these normalized results. We solve the optimization problem for each combination of $c$ and $p$ with $c \in [0, 80]$ with a step of 10 (the step is reduced to 5 in the interval $[0,30]$ to better illustrate smaller values of $c$) and $p \in [0,1]$ with a step of 0.05. We vary a threshold from 0 to the maximum probability score returned by the model, which is 0.78 for the validation month data. 

\begin{figure*} 
    \centering
  \subfloat[Threshold matrix optimized on validation set: $p$ vs. $c$\label{threshold_matrix}]{%
       \includegraphics[width=0.7\textwidth]{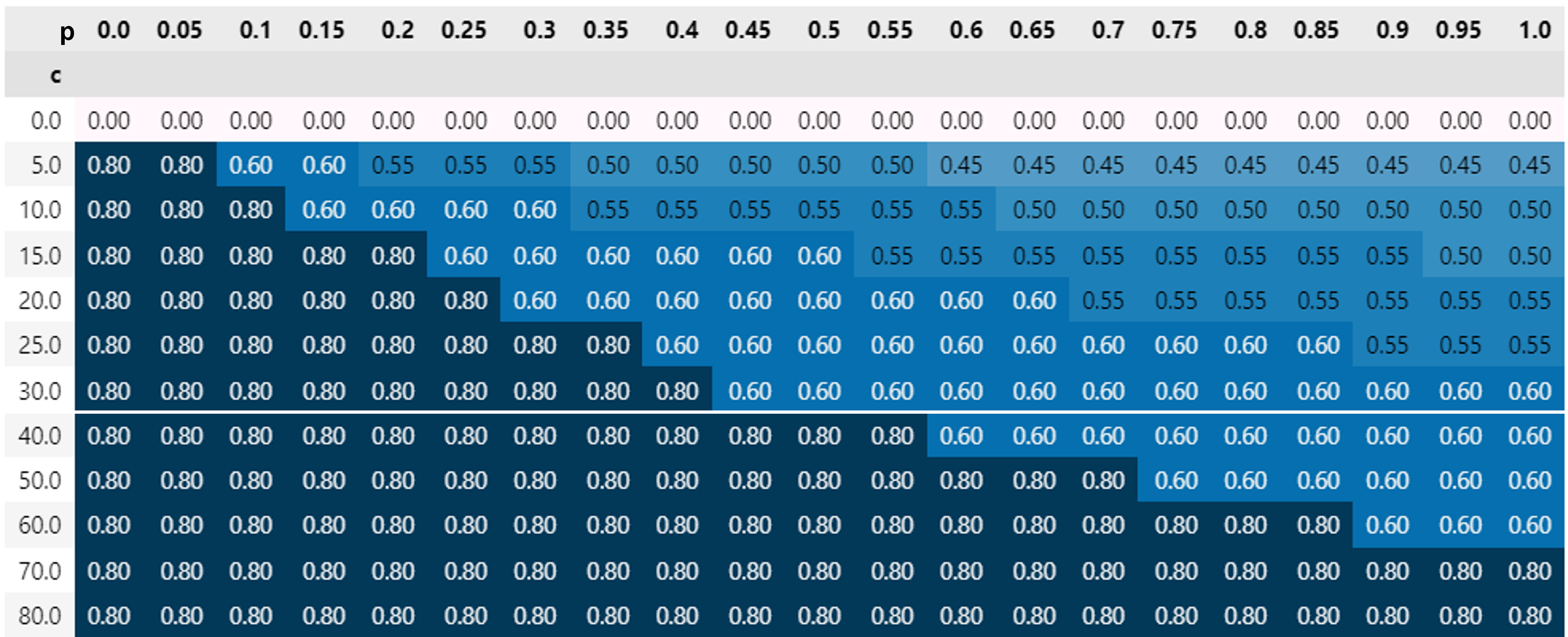}}
       \hspace{0.1cm}
  \subfloat[Profit matrix on test set: $p$ vs. $c$\label{profit_matrix}]{%
        \includegraphics[width=0.8\textwidth]{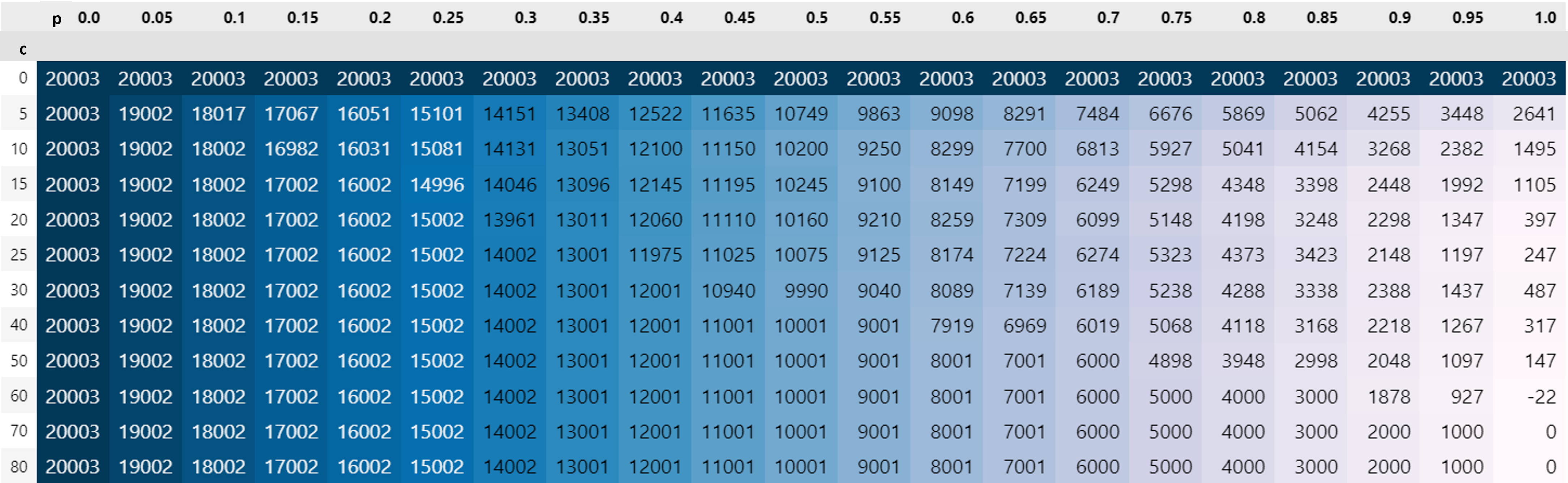}}
  \caption{$p$ vs. $c$ analysis}
  \label{fig:p_vs_c} 
\end{figure*}

Figure \ref{threshold_matrix} shows how the optimal classification threshold changes for different combinations of $c$ and $p$, while Figure \ref{profit_matrix} shows how these thresholds are applied to obtain a profit estimate for these combinations. In particular, when the incentive comes at no cost, the most profitable solution is to contact everyone in the customer base, regardless of the value of $p$. This is also the case when the maximum profit is achieved. This is not surprising, since targeting false positives (\textit{TNI} category) comes at no cost, while targeting true positives (\textit{TI} category) brings non-negative profit. However, this case is not realistic because of the need to contact customers through different channels, which has a non-negative cost. The profit is also maximal when the probability of referral without being targeted is one, i.e., $1-p=1$, which represents a situation where no one is targeted because the profit obtained from the true influencers is higher than when the target campaign is activated, which comes with positive contact costs. 

With an increasing contact cost $c$ and a decreasing probability of referral (i.e., increasing probability $p$ of not referring), the optimal threshold increases: it is no longer economically feasible to contact everyone. This is a more realistic situation, where the model brings added value in selecting the best customers to contact. In this case, the profit becomes smaller with increasing values of probability $p$. This is expected because as probability $p$ increases, the need for the marketing campaign diminishes, as the likelihood of recommendation is high enough without the need to be stimulated by a target. As soon as the contact cost $c$ becomes higher than 60, it is no longer feasible to target customers: the cost becomes higher than the added value of using the model's predictions, regardless of the value of $p$. At this point, the profit is negative and targeted marketing is no longer economically rational. Thus, profit analysis can be used to determine the optimal value of a contact cost at which the marketing campaign still makes sense from a profit perspective. 

\begin{figure*} 
    \centering
  \subfloat[Threshold matrix on validation set: $p$ vs. $ic$\label{threshold_matrix_ic}]{%
       \includegraphics[width=0.8\textwidth]{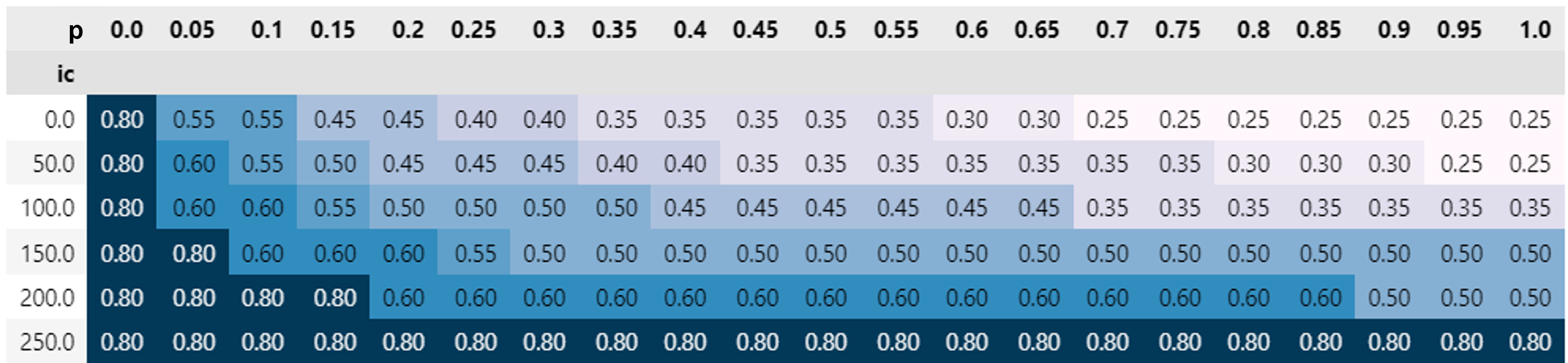}}
       \hspace{0.1cm}
  \subfloat[Profit matrix on test set: $p$ vs. $ic$\label{profit_matrix_ic}]{%
        \includegraphics[width=0.9\textwidth]{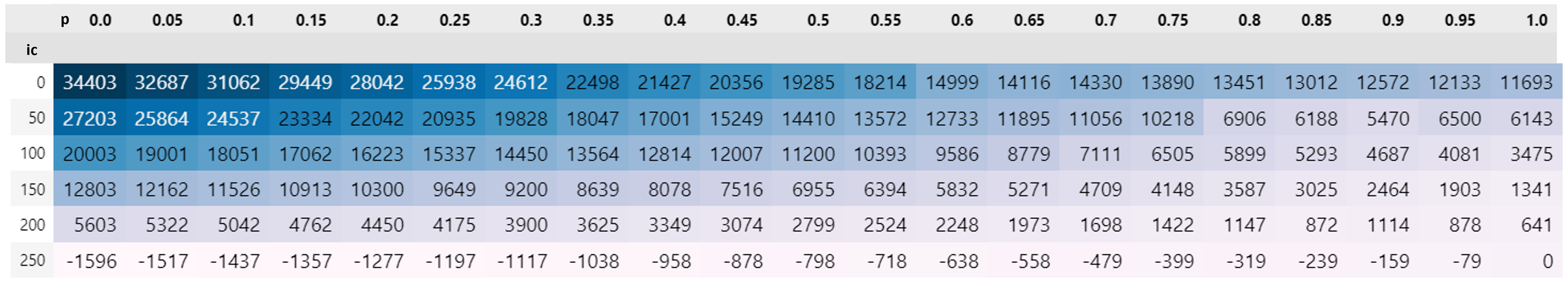}}
 \caption{$p$ vs. $ic$ analysis}
  \label{fig:p_vs_ic} 
\end{figure*}

A similar analysis can be performed for the case where $ic$ is varied instead of $c$ (Figure \ref{fig:p_vs_ic}). In this scenario, we can set $c$ to 3\footnote{The value is chosen based on the conversations with the data provider; it must be chosen based on the use case and the cost structure.} with $ic \in [0, 250]$ with a step of 50 and $p \in [0,1]$ with a step of 0.05. Similar to the $p$ vs. $c$ case, if the probability of referring without being targeted ($1-p$) is 1, the targeting campaign makes no sense: all the influencers will refer to other customers anyway, without being nudged to do so. For increasing probability $p$ and decreasing reward cost $ic$, the optimal threshold value decreases as more customers are predicted as influencers. When the reward cost becomes too high ($\ge$ 200), it is again optimal to not use a model, and rely on referrals without a targeting campaign. When these thresholds are applied to test data to obtain profit estimates, we observe the patterns similar to the $p$ vs. $c$ case: the profit decreases with increasing values of $p$ and increasing reward costs, and becomes negative for reward costs greater than 200. 

The above examples illustrate that prior knowledge is required to deploy influencer prediction models in production. The optimal thresholds should be estimated based on this prior information to get the maximum benefit from the model. The probability $p$, determined by personal brand attitudes and marketing preferences \citep{ryu2007penny}, is beyond the company's control and can be modeled in order to estimate its values depending on aforementioned characteristics. We believe this is an interesting topic for future research. 

\section{Discussion}
\label{discussion}

As illustrated in Section \ref{results}, the GNN-RNN models, in particular the GIN-GRU combination, dominate the baseline models, and this dominance is most pronounced on unseen nodes being crucial for inductive learning that focuses on generalizing to data not encountered by a model. For this configuration, the best combination of performance on seen and unseen nodes is obtained on the largest network of City 3, with consistently high values for all performance metrics. In other words, as task complexity increases, which is the case for inductive learning, the applicability of complex GNN-RNN models is justified. On the contrary, the performance of non-GNN dynamic baseline models, i.e., Features(+PR)-RNN models, deteriorates with increasing network size. While on the smallest network of City 1, these models perform comparable to the GNN-RNN models, the opposite is true for the largest network of City 3. Therefore, for smaller networks, it is advisable to choose simpler models due to the limited performance gain achieved by using GNN-RNN models, but as the networks scale, it is necessary to leverage the complexity of deep learners.

When comparing the two types of GNNs, the GIN encoder emerges as the clear winner in both GNN-RNN and static GNN model types due to its ability to be trained in less time and with fewer $CO_2$ emissions, while consistently outperforming the GAT in terms of detection performance. This superiority can be attributed to the complex network structures we are dealing with: GINs are universal and invariant to the graph structure, whereas GATs are more sensitive to it, which is disadvantageous for complex dynamic networks. This illustrates a trade-off between complexity and performance: the increased complexity of GATs is not justified by a relatively small performance gain. Moreover, in the realm of inductive learning, the importance of the GNN component surpasses that of network dynamics. This is particularly true in the context of the GIN encoder, where models incorporating a GNN encoder consistently exhibit superior AUC and AUPRC values compared to non-GNN dynamic models. Nonetheless, network dynamics play a crucial role and must be captured: the best results are obtained when GNNs and RNNs are coupled as a single model. Given the time and $CO_2$ emissions of the GIN-RNN models, they are also the best candidates for production use: on average, they are the fastest to train with the least carbon footprint. Given the rising cost of energy, this is an important advantage of these models. 

The profit framework highlights the targeting advantages that the proposed models, particularly the GIN-GRU model, can provide. The experiments illustrate that the applicability of the model depends heavily on both the costs of contacting a customer and the value of the reward offered to the referrer. According to \cite{ryu2007penny}, this reward should be non-zero, and the model is found to be useful in cases with lower contact/reward costs and higher probabilities of not referring without being targeted. Since the company can only influence the contact/reward costs, they should be kept as low as possible. However, since there is a positive relationship between a reward size and a referral likelihood \citep{ryu2007penny}, a trade-off between the two should also be considered (see future research directions in Section \ref{conclusion}).

\section{Conclusion}
\label{conclusion}

Network data is getting increased attention due to its ability to reveal complex relationships that may go unnoticed by traditional data sources. Marketing is no exception, as network information can be used to promote products and services through marketing initiatives. Given that DGNNs have demonstrated effective learning from evolving network data \citep{zhu2022learnable}, this paper investigated the optimal strategy for deploying DGNNs to solve the inductive influencer prediction problem, emphasizing profit-sensitive decision-making optimization.
Four DGNN configurations were trained and evaluated on network data from three cities and compared with dynamic non-GNN and static GNN baseline models. The evaluation was performed based on the AUC and AUPRC metrics calculated on seen and unseen nodes, as well as on the time performance and environmental impact of a model. A predict-then-optimize approach was used to illustrate the usability of the models in real business environments. 

First, our study shows that using DGNNs for inductive influencer prediction is highly advantageous. By incorporating dynamic network information, these models demonstrate better generalization to unseen data while maintaining strong detection performance on nodes encountered during training. Among the different types of encoders tested, the GIN model is the most effective due to its ability to handle universal graph structures. We therefore recommend the use of DGNNs with a GIN encoder for influencer prediction tasks that require generalization to new nodes.

Next, we illustrate that applying DGNNs to larger networks provides the best combination of detection performance and model stability. In these cases, DGNNs are particularly useful because they can effectively capture the increasing complexity that cannot be fully captured by centrality metrics such as PageRank. However, for smaller networks with less informative topology, simpler dynamic non-GNN models can perform comparably to DGNNs. Therefore, the choice of model must be carefully evaluated to determine whether the amount of data available justifies the use of more sophisticated models, considering the performance and technical debt of running them.

Finally, the practical application of the model depends heavily on the additional parameters, both controllable and external factors, which can at best be estimated. We conclude that a proper analysis of the cost of the target campaign, together with the cost of referral marketing, should be conducted to find the optimal threshold for final deployment. In our case, we recommend a careful study of the minimum reward necessary to induce a probability of response, using the proposed methods to evaluate the results. A/B testing and similar strategies can be used to fine-tune the value of the incentive.

While this study provides valuable insights into the applicability of DGNNs for influencer prediction, it is important to acknowledge its limitations. The approach used to construct the networks can influence model performance. Investigating the impact of network topology on performance is left for future research. In addition, the profit-driven evaluation framework presented in this study relies on the probability of referring without being targeted, which requires explicit modeling. Future research can focus on modeling this probability.

\section*{Acknowledgments}
The research was sponsored by the ING Chair on Applying Deep Learning on Metadata as a Competitive Accelerator. The last author acknowledges the support of the NSERC Discovery Grant program [RGPIN-2020-07114]. This research was undertaken, in part, thanks to funding from the Canada Research Chairs program [CRC-2018-00082]. This work was enabled in part by support provided by Compute Ontario (\url{computeontario.ca}), Calcul Québec (\url{calculquebec.ca}), and the Digital Research Alliance of Canada (\url{alliancecan.ca}). The third author acknowledges the support of the Icelandic Research Fund (IRF) [grant number 228511-051]. We thank the anonymous company that provided the data.

\appendices
\section{Node features}
\label{appen:node_features}

\begin{table}[H]
\fontsize{6}{8}\selectfont 
  \caption{Node features description}
  \label{node_features}
  \centering
  \begin{tabular}{p{3.5in}p{0.6in}p{0.3in}}
    \toprule                 \\
    \textbf{Feature group description} & \textbf{Type} & \textbf{\#feat.} \\
    \midrule
    \# no payments from the contract creation date to the payment date & Numerical  & 1  \\
    Outstanding balance of a credit card at a cut-off date & Numerical  & 1  \\
    Payments between: \newline 1) start date  of the billing period and cut-off date (a final date of the billing period); \newline 2) cut-off date to payment date (the date when a customer is supposed to pay) & Numerical  & 2  \\
    Credit card consumption in delivery app (total, average and standard deviation) between: \newline 1) start date to cut-off date; \newline 2) cut-off date to payment date & Numerical  & 6 \\
     Credit card consumption in other commercials (total, average and standard deviation) between: \newline 1) start date to cut-off date; \newline 2) cut-off date to payment date & Numerical  & 6 \\
     Number of transactions between: \newline 1) start date to cut-off date; \newline 2) cut-off date to payment date & Numerical & 2 \\
     Credit limit & Numerical & 1 \\
     Number of no payments between the contract creation date and the payment date & Numerical & 1 \\
     Bureau score of the user at an application date & Numerical & 1 \\
     Estimated income at an application date & Numerical & 1 \\
     Amount of increase in the credit limit & Numerical & 1 \\
     User's operative state at the payment date & Categorical & 1 \\
     Credit limit increase indicator & Categorical & 1 \\
     Counter of past referrals & Numerical & 1 \\
    \bottomrule
  \end{tabular}
\end{table}

\section{Network characteristics}
\label{appen:networks}
\begin{figure}[H]
    \centering
  \subfloat[City 1\label{fig:city1}]{%
       \includegraphics[width=0.5\linewidth]{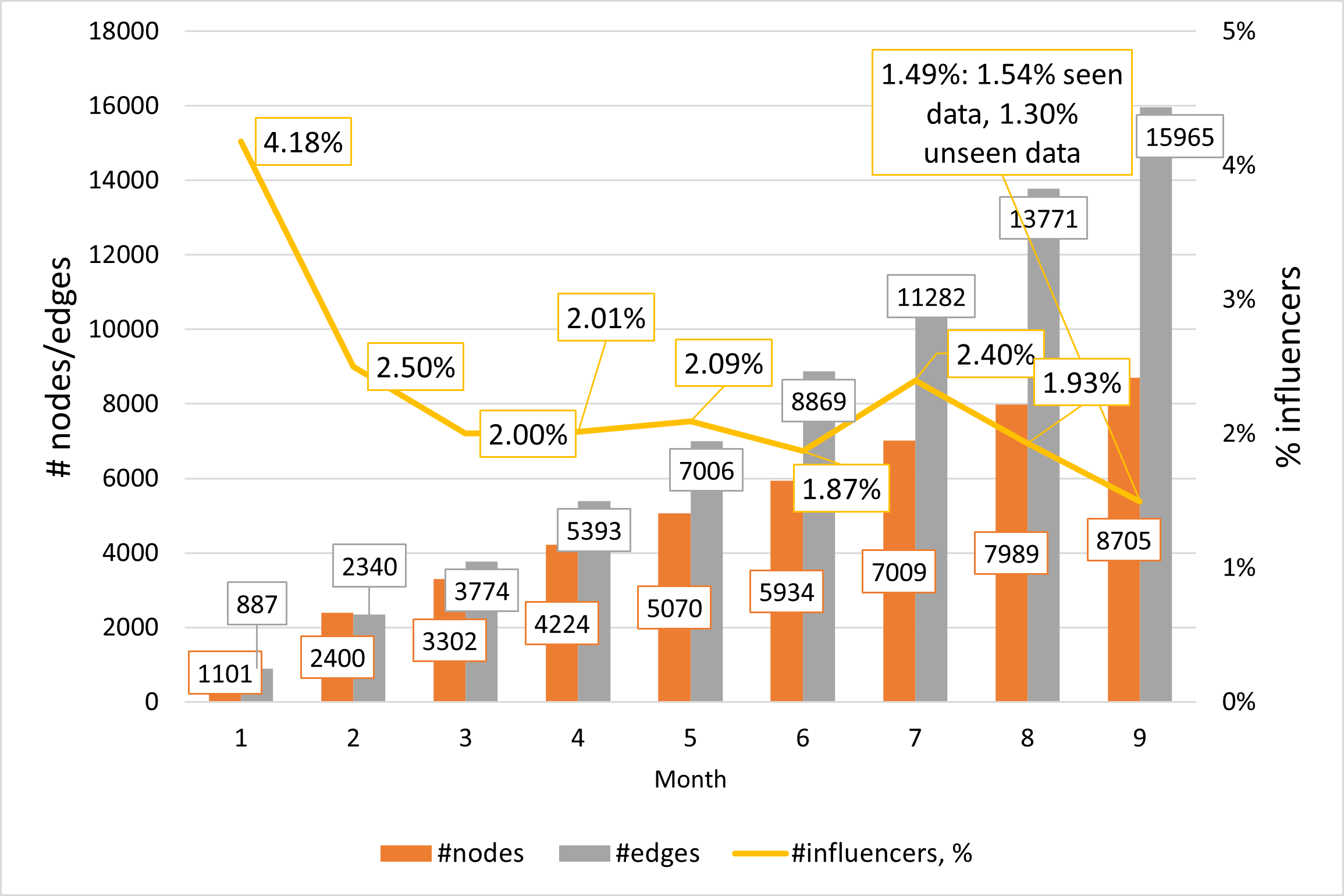}} \\
  \subfloat[City 2\label{fig:city2}]{%
        \includegraphics[width=0.5\linewidth]{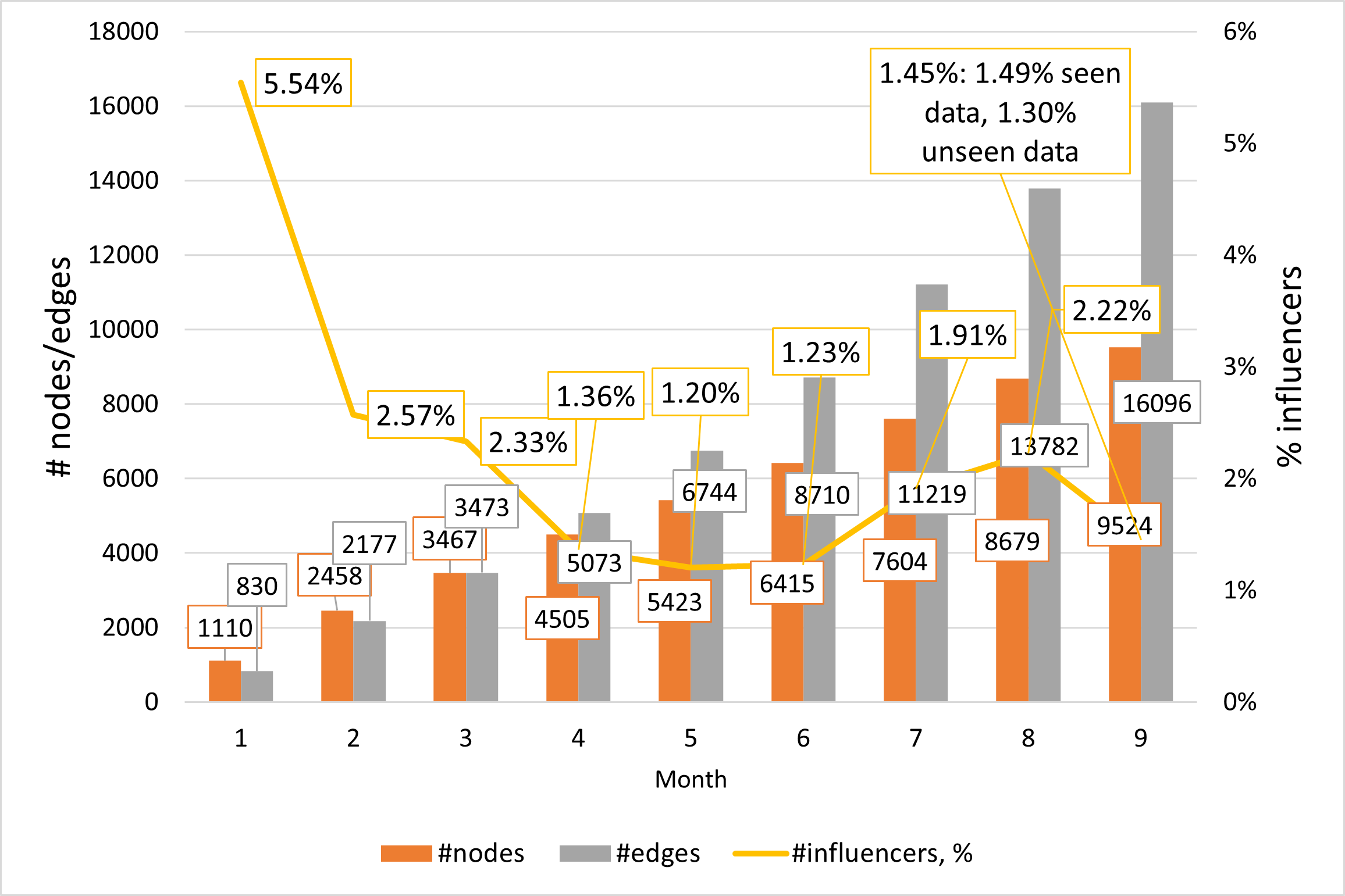}} \\
\subfloat[City 3\label{fig:city3}]{%
        \includegraphics[width=0.5\linewidth]{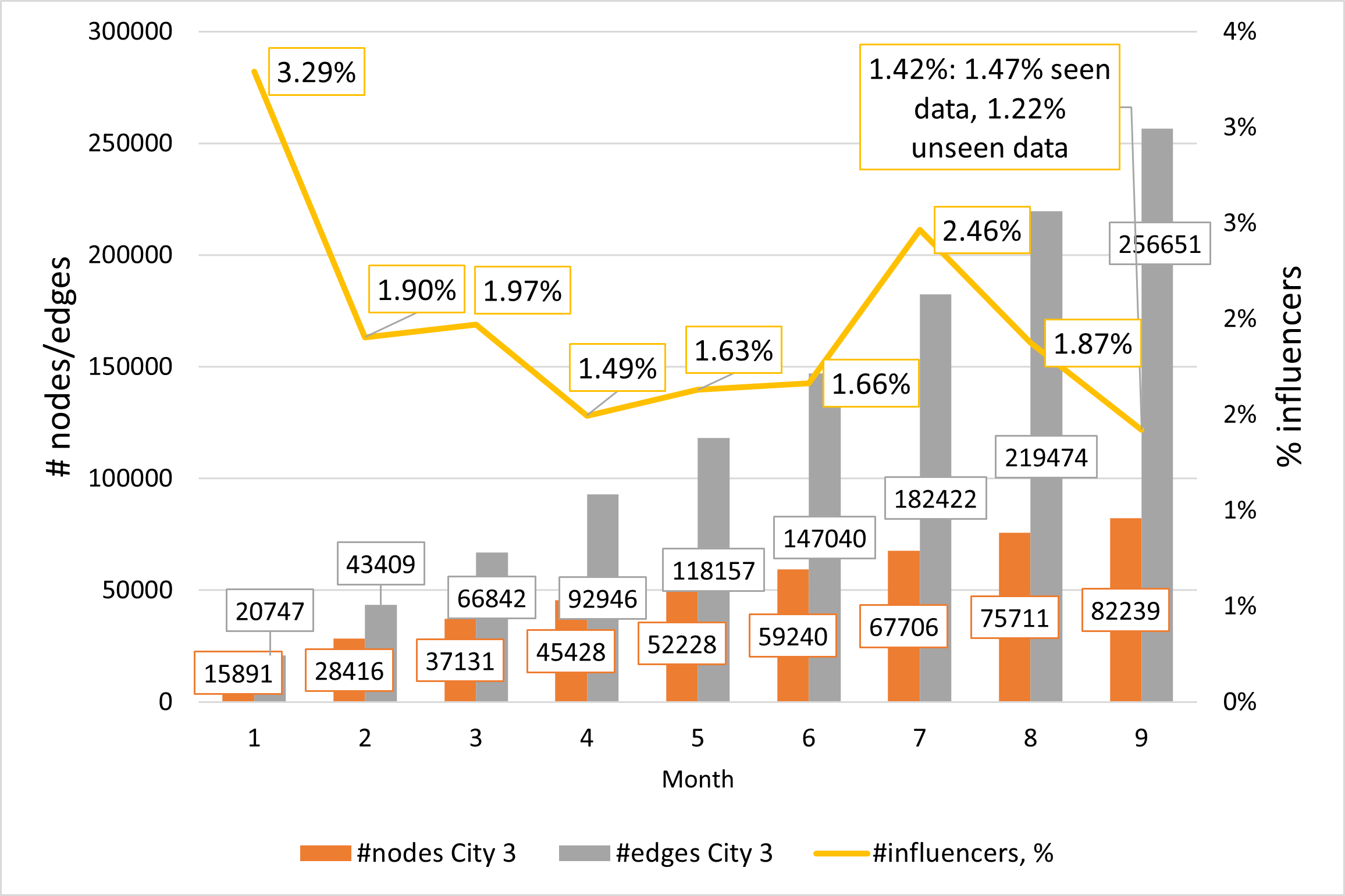}}
  \caption{Network characteristics}
  \label{fig:networs_char} 
\end{figure}

\section{Hyperparameter configurations}
\label{app:model_configurations}

\begin{table}[H]
\caption{Model configurations: City 1}
\fontsize{6}{8}\selectfont 
\centering
\begin{tabular}{@{}p{3cm}p{0.9cm}p{0.9cm}p{0.9cm}p{0.9cm}p{0.9cm}p{0.9cm}p{0.9cm}@{}}
\hline
\textbf{Model} & \textbf{\#layers GNN} & \textbf{\#emb. GNN} & \textbf{\#hid.dim. GNN} & \textbf{SMOTE rate} & \textbf{\#layers RNN} & \textbf{\#hid.dim. RNN} & \textbf{Loss} \\ \hline
GIN-LSTM & 2 & 512 & 512 & 0 & 2 & 256 & Focal \\
GIN-GRU & 2 & 512 & 256 & 0 & 2 & 256 & BCE \\
GAT-LSTM & 4 & 256 & 512 & 0 & 1 & 256 & Focal \\
GAT-GRU & 4 & 256 & 512 & 0 & 1 & 256 & BCE \\
Features+PR-LSTM & / & / & / & 0 & 1 & 512 & Focal \\
Features+PR-GRU & / & / & / & 0 & 1 & 512 & Focal \\
Features-LSTM & / & / & / & 0.5 & 1 & 256 & Focal \\
Features-GRU & / & / & / & 0 & 1 & 512 & Focal \\
Static GAT & 4 & 512 & 256 & 0.5 & / & / & BCE \\
Static GIN & 2 & 256 & 512 & 0.5 & / & / & Focal \\ \hline
\end{tabular}
\label{tab:model_configurations_city1}
\end{table}

\begin{table}[H]
\caption{Model configurations: City 2}
\fontsize{6}{8}\selectfont 
\centering
\begin{tabular}{@{}p{3cm}p{0.9cm}p{0.9cm}p{0.9cm}p{0.9cm}p{0.9cm}p{0.9cm}p{0.9cm}@{}}
\hline
\textbf{Model} & \textbf{\#layers GNN} & \textbf{\#emb. GNN} & \textbf{\#hid.dim. GNN} & \textbf{SMOTE rate} & \textbf{\#layers RNN} & \textbf{\#hid.dim. RNN} & \textbf{Loss} \\ \hline
GIN-LSTM & 2 & 512 & 256 & 0 & 1 & 256 & Focal \\
GIN-GRU & 2 & 512 & 512 & 0 & 2 & 256 & Focal \\
GAT-LSTM & 2 & 256 & 512 & 0 & 2 & 512 & Focal \\
GAT-GRU & 4 & 256 & 512 & 0 & 2 & 256 & Focal \\
Features+PR-LSTM & / & / & / & 0 & 1 & 512 & Focal \\
Features+PR-GRU & / & / & / & 0 & 2 & 256 & Focal \\
Features-LSTM & / & / & / & 0 & 1 & 512 & Focal \\
Features-GRU & / & / & / & 0.5 & 1 & 256 & Focal \\
Static GAT & 4 & 256 & 256 & 0 & / & / & BCE \\
Static GIN & 2 & 512 & 256 & 0 & / & / & Focal \\ \hline
\end{tabular}
\label{tab:model_configurations_city2}
\end{table}

\begin{table}[H]
\caption{Model configurations: City 3}
\fontsize{6}{8}\selectfont 
\centering
\begin{tabular}{@{}p{3cm}p{0.9cm}p{0.9cm}p{0.9cm}p{0.9cm}p{0.9cm}p{0.9cm}p{0.9cm}@{}}
\hline
\textbf{Model} & \textbf{\#layers GNN} & \textbf{\#emb. GNN} & \textbf{\#hid.dim. GNN} & \textbf{SMOTE rate} & \textbf{\#layers RNN} & \textbf{\#hid.dim. RNN} & \textbf{Loss} \\ \hline
GIN-LSTM & 2 & 512 & 512 & 0 & 1 & 256 & Focal \\
GIN-GRU & 2 & 512 & 512 & 0 & 1 & 512 & Focal \\
GAT-LSTM & 2 & 256 & 256 & 0 & 1 & 512 & Focal \\
GAT-GRU & 2 & 256 & 256 & 0 & 2 & 256 & Focal \\
Features+PR-LSTM & / & / & / & 0 & 1 & 512 & BCE \\
Features+PR-GRU & / & / & / & 0.5 & 1 & 512 & Focal \\
Features-LSTM & / & / & / & 0.5 & 1 & 256 & Focal \\
Features-GRU & / & / & / & 0 & 1 & 512 & BCE \\
Static GAT & 4 & 512 & 256 & 0.5 & / & / & Focal \\
Static GIN & 2 & 512 & 512 & 0.5 & / & / & BCE \\ \hline
\end{tabular}
\label{tab:model_configurations_city3}
\end{table}

\bibliographystyle{apacite}
\bibliography{bibliography.bib}

\end{document}